\newcommand{\fig}[1]{\mbox{Fig.\hspace{0.2em}\ref{#1}}}
\newcommand{\eqn}[1]{\mbox{Eq.\hspace{0.2em}\ref{#1}}}
\newcommand{\sect}[1]{\mbox{\S\ref{#1}}}
\newcommand{\tbl}[1]{\mbox{Table\hspace{0.3em}\ref{#1}}}

\newcommand{\fij}{f_{ij}}

\newcommand{\rij}{r_{ij}}
\newcommand{\sij}{s_{ij}}
\newcommand{\vt}{v_\mathrm{t}}
\newcommand{\NP}{N_\mathrm{p}}
\newcommand{\NT}{T}
\newcommand{\Nh}{N_\mathrm{hyp}}
\newcommand{\Ng}{N_\mathrm{g}}

\newcommand{\KK}[3]{K(#1; #2, #3)}
\newcommand{\SSS}[2]{S(#1; #2)}

\newcommand{\dmax}{D_\mathrm{max}}
\newcommand{\uC}{u_\mathrm{c}}
\newcommand{\uH}{u_\mathrm{h}}
\newcommand{\vC}{v_\mathrm{c}}
\newcommand{\vH}{v_\mathrm{h}}
\newcommand{\wC}{w_\mathrm{c}}
\newcommand{\wH}{w_\mathrm{h}}
\newcommand{\xC}{x_\mathrm{c}}
\newcommand{\xH}{x_\mathrm{h}}

\def\P{{\mathbb P}}
\def\E{{\mathbb E}}
\def\1{{\mathbf 1}}

\documentclass[apj]{emulateapj}
\usepackage{graphicx}
\usepackage{natbib}
\usepackage{amssymb}
\usepackage{natbib}

\begin{document}

\title{The TAOS Project: Statistical Analysis of Multi-Telescope Time Series
Data}
\author{
M.~J.~Lehner\altaffilmark{1,2,3},
N.~K.~Coehlo\altaffilmark{4},
Z.-W.~Zhang\altaffilmark{1,5},
F.~B.~Bianco\altaffilmark{6,7,2,3},
J.-H.~Wang\altaffilmark{1,5},
J.~A.~Rice\altaffilmark{4},
P.~Protopapas\altaffilmark{8,3},
C.~Alcock\altaffilmark{3},
T.~Axelrod\altaffilmark{9},
Y.-I.~Byun\altaffilmark{10},
W.~P.~Chen\altaffilmark{5},
K.~H.~Cook\altaffilmark{11},
I.~de~Pater\altaffilmark{12},
D.-W.~Kim\altaffilmark{10,3,8},
S.-K.~King\altaffilmark{1},
T.~Lee\altaffilmark{1},
S.~L.~Marshall\altaffilmark{13,11},
M.~E.~Schwamb\altaffilmark{14},
S.-Y.~Wang\altaffilmark{1} and
C.-Y.~Wen\altaffilmark{1}}
\altaffiltext{1}{Institute of Astronomy and Astrophysics, Academia Sinica.
 P.O. Box 23-141, Taipei 106, Taiwan}
\email{mlehner@asiaa.sinica.edu.tw}
\altaffiltext{2}{Department of Physics and Astronomy, University of
 Pennsylvania, 209 South 33rd Street, Philadelphia, PA 19104}
\altaffiltext{3}{Harvard-Smithsonian Center for Astrophysics, 60 Garden Street,
 Cambridge, MA 02138}
\altaffiltext{4}{Department of Statistics, University of California Berkeley,
 367 Evans Hall, Berkeley, CA 94720}
\altaffiltext{5}{Institute of Astronomy, National Central University, No. 300,
 Jhongda Rd, Jhongli City, Taoyuan County 320, Taiwan}
\altaffiltext{6}{Department of Physics, University of California Santa Barbara, 
Mail Code 9530,  Santa Barbara CA 93106-9530 }
\altaffiltext{7}{Las Cumbres Observatory Global Telescope Network, Inc.
6740 Cortona Dr. Suite 102, Santa Barbara, CA 93117}
\altaffiltext{8}{Initiative in Innovative Computing, Harvard University,
 60 Oxford St, Cambridge, MA 02138}
\altaffiltext{9}{Steward Observatory, 933 North Cherry Avenue, Room N204
 Tucson AZ 85721}
\altaffiltext{10}{Department of Astronomy, Yonsei University, 134 Shinchon,
 Seoul 120-749, Korea}
\altaffiltext{11}{Institute of Geophysics and Planetary Physics, Lawrence
 Livermore National Laboratory, Livermore, CA 94550}
\altaffiltext{12}{Department of Astronomy, University of California Berkeley,
 601 Campbell Hall, Berkeley CA 94720}
\altaffiltext{13}{Kavli Institute for Particle Astrophysics and Cosmology,
 2575 Sand Hill Road, MS 29, Menlo Park, CA 94025}
\altaffiltext{14}{Division of Geological and Planetary Sciences,
 California Institute of Technology, 1201 E. California Blvd., Pasadena, CA
 91125}

\begin{abstract}
The Taiwanese-American Occultation Survey (TAOS) monitors fields of up
to $\sim$1000~stars at 5~Hz simultaneously with four small telescopes
to detect occultation events from small ($\sim$1~km) Kuiper Belt
Objects (KBOs). The survey presents a number of challenges, in
particular the fact that the occultation events we are searching for
are extremely rare and are typically manifested as slight flux drops
for only one or two consecutive time series measurements. We have
developed a statistical analysis technique to search the
multi-telescope data set for simultaneous flux drops which provides a
robust false positive rejection and calculation of event
significance. In this paper, we describe in detail this statistical
technique and its application to the TAOS data set.
\end{abstract}

\keywords{Solar System, Astronomical Techniques, Data Analysis and Techniques}

\section{Introduction}
\label{sec:intro}
The Taiwanese-American Occultation Survey operates four small
telescopes \citep{fed, 2009AJ....138.1893W, 2008ApJ...685L.157Z,
  2009PASP..121..138L} at Lulin Observatory in central Taiwan to
search for occultations by small ($\sim$1~km diameter) KBOs
\citep{2009Natur.462..895S, 2009arXiv0910.5598W, 2009AJ....138..568B,
  2009AJ....137.4270B, 2008AJ....135.1039B, 2007AJ....134.1596N,
  2007MNRAS.378.1287C, 2006AJ....132..819R, cooray,
  2003ApJ...587L.125C, 2003ApJ...594L..63R}. Such occultation events
are extremely rare (estimated rates range from $10^{-4}$~to
$10^{-2}$~events per star per year), they are very short in duration
($\lesssim$200~ms), and at the 5~Hz observing cadence used by TAOS,
they result in measured flux drops of typically $\lesssim$30\% in one
or two consecutive points. This presents a number of challenges, in
particular the identification of false positive events of statistical
origin and candidate events which are in fact of terrestrial origin
(e.g. birds, airplanes, and extreme scintillation events). We reject
these false positive events by requiring simultaneous detection in
multiple telescopes.

A second challenge is finding a robust method to determine the
statistical significance of any candidate events. The noise
distribution of each lightcurve is not known \emph{a priori}, due to
non-Poisson and non-Gaussian processes on the tails of the flux
distributions. The typical stars in our fields have magnitudes $R\sim
13$ and a signal to noise ratio (SNR) of $\sim$10. Moreover, TAOS
monitors fields for durations of up to 1.5~hours, and changes in
atmospheric transparency and airmass introduce further uncertainties
into the flux measurements.

To overcome these difficulties, we have developed non-parametric
techniques using \emph{rank statistics.} Rank statistics facilitate a
simultaneous analysis of multi-telescope photometric measurements to
enable a robust determination of event significance and false positive
rejection, which are independent of the underlying noise distributions
of the lightcurves being analyzed. Rank statistics and their
application to the TAOS lightcurves will be described in the following
sections. In \sect{sec:occ} we review the characteristics of
occultation events and describe how such events would appear in the
data. In \sect{sec:ranks} we discuss the rank product statistical test
used to calculate event significance and the false positive rate. In
\sect{sec:filter} we describe the lightcurve filtering techniques and
diagnostic tests used to ensure that the rank product statistical test
is valid, and in \sect{sec:newtests} we describe a new and more robust
set of diagnostics tests.

The following definitions apply throughout the remainder of this
paper.  We define a \emph{data run} as a consecutive series of
multi-telescope observations of a given star field made at a cadence
of 5~Hz. Typical data runs last 1.5~hours, comprising 27,000 time
series images on each telescope. We define a \emph{lightcurve set} as
a set of multi-telescope lightcurves of a single star during a given
data run. There are typically 300--500 stars in an image, and hence
300--500 lightcurve sets in a data run. We adopt the standard
statistical notation wherein we denote a random variable with an upper
case letter, and use the corresponding lower case letter for an actual
value for that variable (e.g. $Z$ is a random variable which could
take on a value of $z$). We use the function $p()$ to describe a
probability density distribution, and $\P()$ to describe an actual
probability. Finally, we note that the four telescopes are labeled
TAOS~A, TAOS~B, TAOS~C, and TAOS~D. TAOS~C came online in August of
2008, and to date no data from this telescope has been analyzed. All
example lightcurves shown in this paper come from telescopes A, B and
D.

\begin{figure*}
\plottwo{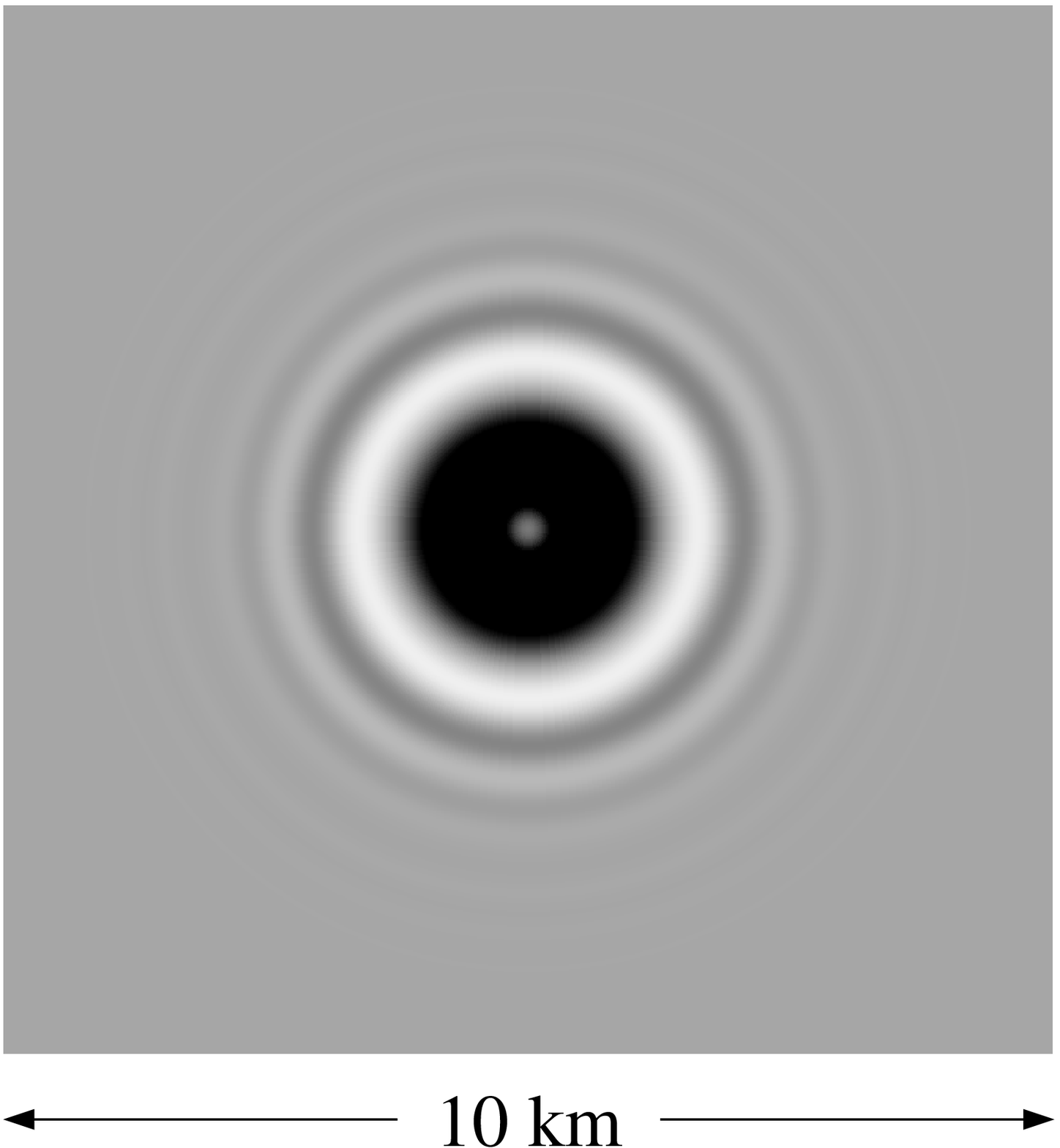}{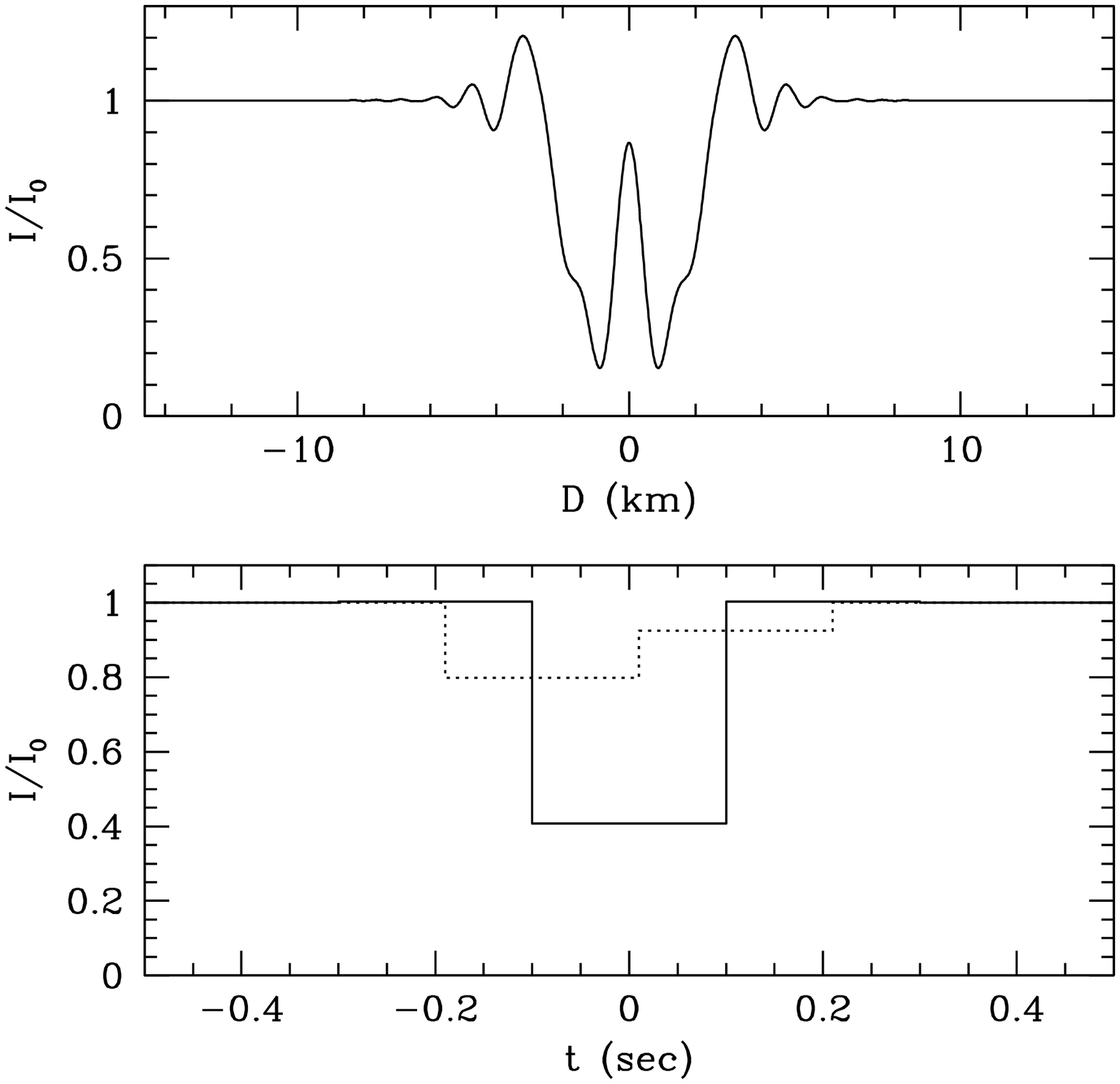}
\caption[]{Left panel: diffraction shadow projected onto the surface
  of the Earth from a 3~km diameter KBO at 43 AU. Right panel top:
  perfectly sampled lightcurve assuming zero impact parameter. Right
  panel bottom: same lightcurve as sampled by the TAOS system at
  5~Hz. Solid curve has measurements centered on event, dotted line
  shows lightcurve where sampling is out of phase with event.}
\label{fig:diff}
\end{figure*}

\section{Occultations by Kuiper Belt Objects}
\label{sec:occ}
An occultation event occurs when an object passes between the
telescope and a distant star \citep{2009AJ....137.4270B,
  2007AJ....134.1596N, 2003ApJ...594L..63R}. The Earth and the
occulting object are in relative motion, inducing a variation in the
measured stellar flux over time. The target population for TAOS is
small ($\sim$1~km diameter) KBOs, whose sizes are on the order of the
\emph{Fresnel scale}, which is given by
\begin{displaymath}
F = \sqrt{\frac{\lambda\Delta}{2}},
\end{displaymath}
where $\lambda$ is the wavelength of observation and $\Delta$ is the
observer--KBO distance. For TAOS, the median wavelength of observation
is $\lambda \approx 600$~nm, and the typical distance to KBOs is
43~AU, resulting in $F = 1.4$~km. Occultation events by KBOs with
diameters $D \lesssim 10$~km thus show significant diffraction
effects. This is illustrated in the left panel of \fig{fig:diff},
which shows a simulated occultation ``shadow'' from a 3~km diameter
KBO projected onto the surface of the Earth.

The timescale of an occultation event is set by the relative velocity
between the KBO and observer, the size of the occultation shadow, and
the impact parameter (minimum distance between the KBO and the line of
sight to the target star). The relative velocity between the Earth and
KBO is given by
\begin{equation}
  v_\mathrm{rel} = v_\mathrm{E}\left[\cos\phi -
  \left(\frac{1\,\mathrm{AU}}{\Delta}\right)^{\frac{1}{2}}
    \left(1-\frac{1\,\mathrm{AU}^2}{\Delta^2}\sin^2\phi\right)^{\frac{1}{2}}
    \right],
\label{eq:vrel}
\end{equation}
where $\phi$ is the angle of observation between the occulted star and
opposition, and $v_\mathrm{E} = 29.8$~km\,s$^{-1}$ is the velocity of
the Earth around the Sun. The event width (the length of the chord
across the occultation shadow where it crosses the telescope) is given
by
\begin{displaymath}
W = \sqrt{H^2 - b^2},
\end{displaymath}
where $b$ is the impact parameter, and $H$ is the event cross
section, which we define as the diameter of the first Airy ring of the
diffraction shadow, and which can be approximated by
\citep{2007AJ....134.1596N}
\begin{equation}
H \approx \left[(2\sqrt{3}F)^{\frac{3}{2}} + D^{\frac{3}{2}}
  \right]^{\frac{2}{3}} + \theta_*\Delta,
\label{eq:H}
\end{equation}
where $\theta_*$ is the angular size of the occulted star.

For the very small objects ($D \lesssim 1$~km) targeted by this survey
and stars with small angular diameters (the vast majority of stars
covered by this survey), the minimum event cross section is set by the
Fresnel scale:
\begin{equation}
H_\mathrm{min} \approx 2\sqrt{3}F.
\label{eq:hmin}
\end{equation}
At 43~AU, $H_\mathrm{min} \approx 5$~km. At opposition ($\phi = 0$),
$v_\mathrm{rel} \approx 25$~km\,s$^{-1}$, and with $b = 0$ the
resulting event duration is 200~ms, with the duration getting smaller
as $b$ is increased.

This is illustrated in the right panels of \fig{fig:diff}. The top
panel shows a slice through the simulated diffraction shadow, assuming
the KBO crosses the line of sight to the star ($b = 0$). Note the
event width, given by the distance between the two top peaks, is about
5~km. (In this case, the event width is dominated by the Fresnel
scale, so the approximation given in \eqn{eq:hmin} applies.) The
bottom panel shows this event as it would be measured by the TAOS
system at 5~Hz. The solid line shows the lightcurve which would be
measured if the sampling was in phase with the event, that is, the
measurement at $t = 0$ is centered on the epoch when the KBO is
centered on the line of sight to the target star. The dotted line
shows the same event with the sampling out of phase with the event.

\begin{figure}
\plotone{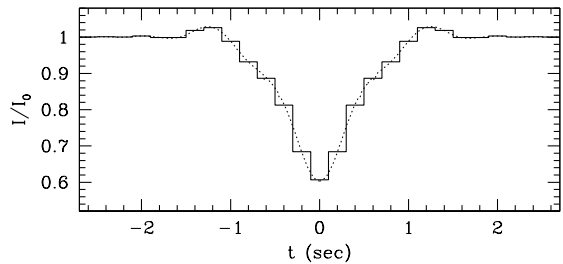}
\caption[]{Simulated lightcurve with an occultation by a 5~km object
  at 500~AU, measured at $70^\circ$ from opposition. Diffraction
  features are smoothed out due to finite angular size of the occulted
  star. Dotted line is the infinitesimally sampled lightcurve, and the
  solid line indicates the lightcurve as would be measured with 5~Hz
  sampling. See \citet{2007AJ....134.1596N} for a discussion of
  occultation events from objects at such distances.}
\label{fig:sdiff}
\end{figure}

Typical occultation events for small KBOs at opposition will thus
manifest themselves in the TAOS data as a reduction in flux on one or
two consecutive photometric measurements of a star with an otherwise
flat lightcurve. However, when observing away from opposition, the
relative velocity decreases, as indicated by
\eqn{eq:vrel}. Furthermore, TAOS is also sensitive to objects more
distant than the Kuiper Belt. The discovery of Sedna
\citep{2004ApJ...617..645B} indicates the possibility of a large,
heretofore unknown population of objects at distances of 100~to
1000~AU \citep[see][and references therein]{2009AJ....138.1893W}. Such
events will also be of a longer duration due to the increased angular
size of the Fresnel scale, as indicated by \eqn{eq:H}. \fig{fig:sdiff}
shows a simulated lightcurve with an occultation by a 5~km object at
500~AU, observed at $70^\circ$ from opposition. The width of the event
is about 22~km, and with a relative velocity of about 9~km\,s$^{-1}$,
the event duration is about 2.5~seconds, corresponding to a total of
13~measurements at our cadence of 5~Hz. (Once again, the approximation
for the event width given in \eqn{eq:hmin} applies.)

The goals of the TAOS statistical analysis described in this paper are
to find as many such events as possible, minimize the false positive
rate, and provide a method to robustly estimate the statistical
significance of any candidate event. The statistical technique should
be sensitive to both the one and two point events shown in
\fig{fig:diff} and the longer duration events such as that shown in
\fig{fig:sdiff}. In the following sections the application of rank
statistics to meet these goals will be described. The discussion will
begin with a focus on single point events, and the extension of the
statistical technique to multi-point observations will be presented in
\sect{sec:multipoint}.

\section{Rank Statistics}
\label{sec:ranks}

The idea of rank statistics is quite simple, and is best introduced
with a single series. Take a time series of flux measurements $f_1,
\ldots, f_{\NP}$ from one telescope, where $\NP$ is the number of
points in the time series. Replace each flux measurement $f_j$ with
its rank $r_j$.  That is, the lowest $f_j$ will be assigned rank 1 and
the highest assigned rank $\NP$. If we use a total of
$\NT$~telescopes, we replace the time series for each telescope with
its rank within the lightcurve from that telescope, giving a set of
$\NT$ rank time series $\rij$. Thus for each time point $t_j$, we have
a \emph{rank tuple}
\begin{displaymath}
(r_{1j}, \ldots, r_{\NT j}).
\end{displaymath}

If, for each telescope $i$, the rank $\rij$ follows a uniform
distribution on $\{1, \ldots, \NP\}$ at each time point $t_j$, and if
the lightcurves $\fij$ are independent between the different
telescopes, then each rank tuple combination is equally likely at each
time point, and we can calculate exact probability distribution of
these rank tuples. The calculation of the probability distribution of
the raw data is impossible to perform on the original time series
measurements, since the underlying distributions of the flux
measurements $\fij$ in each lightcurve are unknown. That is, by
working with the ranks, we replace something unknown, the distribution
of the data, with something known, the distribution of the ranks.

A time series is \textit{stationary} if the distribution of any finite
subset of the series is invariant under time shift. A stationary time
series $f_j$ is \emph{ergodic in mean} if, for any function $G$ with
an expected value $\E(G(f)) < \infty$, we have the following
convergence with probability of 1 (law of large numbers):
\begin{displaymath}
\lim_{n \rightarrow \infty} \frac{1}{n} \sum_{i=1}^n G(f_i) \rightarrow \E(G(f)).
\end{displaymath}
It can be shown that if a time series $f_j$ of length $N$ is ergodic
in mean, then the distribution of ranks $r_{j} / N$ will converge to
the uniform distribution for any sequence $1 \le j \le N$, and it is
well known that ergodicity in mean can be assured under very weak
assumptions on temporal dependence within a lightcurve. This proof is
beyond the scope of this paper, but it is published in \citet{nate}.

Therefore, if the data $f_j$ are stationary and ergodic in mean, this
implies that at each time point $t_j$, $r_j$ will be uniform on $\{1,
\ldots, \NP\}$. In addition, if the lightcurves from different
telescopes are independent, then the rank tuples $(r_{1j}, \ldots,
r_{\NT j})$ will be uniform on
\begin{displaymath}
  \{1,2,...,\NP\}^{\NT},
\end{displaymath}
and we can calculate exact probability distributions of the rank
tuples. However, most of the lightcurve sets exhibit slowly varying
trends that are highly correlated between the different telescopes, so
our lightcurves are neither uncorrelated nor stationary. We have
developed a filtering algorithm to remove these trends, and in most
cases the resulting individual lightcurves can be plausibly modeled as
stationary and ergodic in mean. In most cases the correlations between
the lightcurves from different telescopes are also removed by the
application of the filter. This filtering algorithm will be described
in \sect{sec:filter}.

\begin{figure}
\plotone{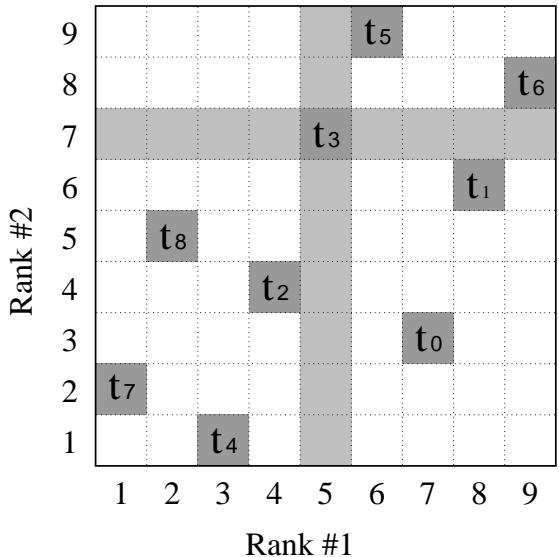}
\caption[]{Schematic of a rank-rank diagram, with $\NP=9$. Axes are
  ranks of photometric intensity for individual data points on two
  different telescopes. A single two-telescope photometric measurement
  will correspond to a rank doublet on this plot. These are marked
  with the dark squares and labeled with the time at which they where
  measured. For example, note the highlighted rank pair at (5,7),
  measured at time $t_3$. Note that each rank value occurs once and
  only once in the time series for each telescope.}
\label{fig:rr}
\end{figure}

\fig{fig:rr} introduces the \emph{rank-rank diagram}, which is a
scatter-plot of the ranks on two telescopes. Similar plots will be
used throughout the remainder of the paper to illustrate various
statistical tests. Note that each rank must occur once and only once
in each time series. Thus there must be exactly one point in each row
and in each column. The ranks within a single lightcurve are thus not
independently distributed. However, if the conditions on ergodicity in
mean and no dependence between telescopes are met, then the rank pairs
will be uniformly distributed throughout the diagram.

\begin{figure*}
\plottwo{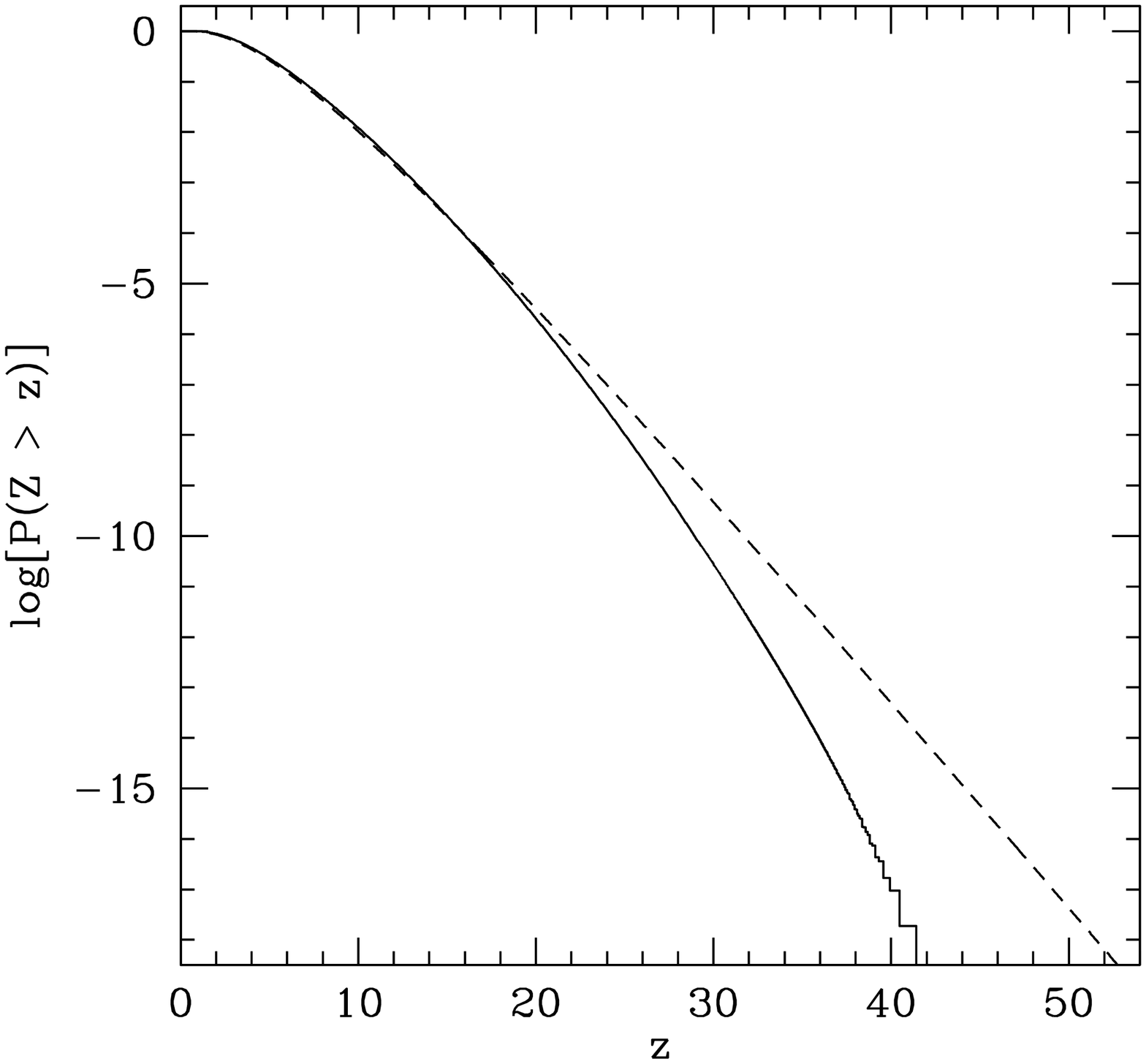}{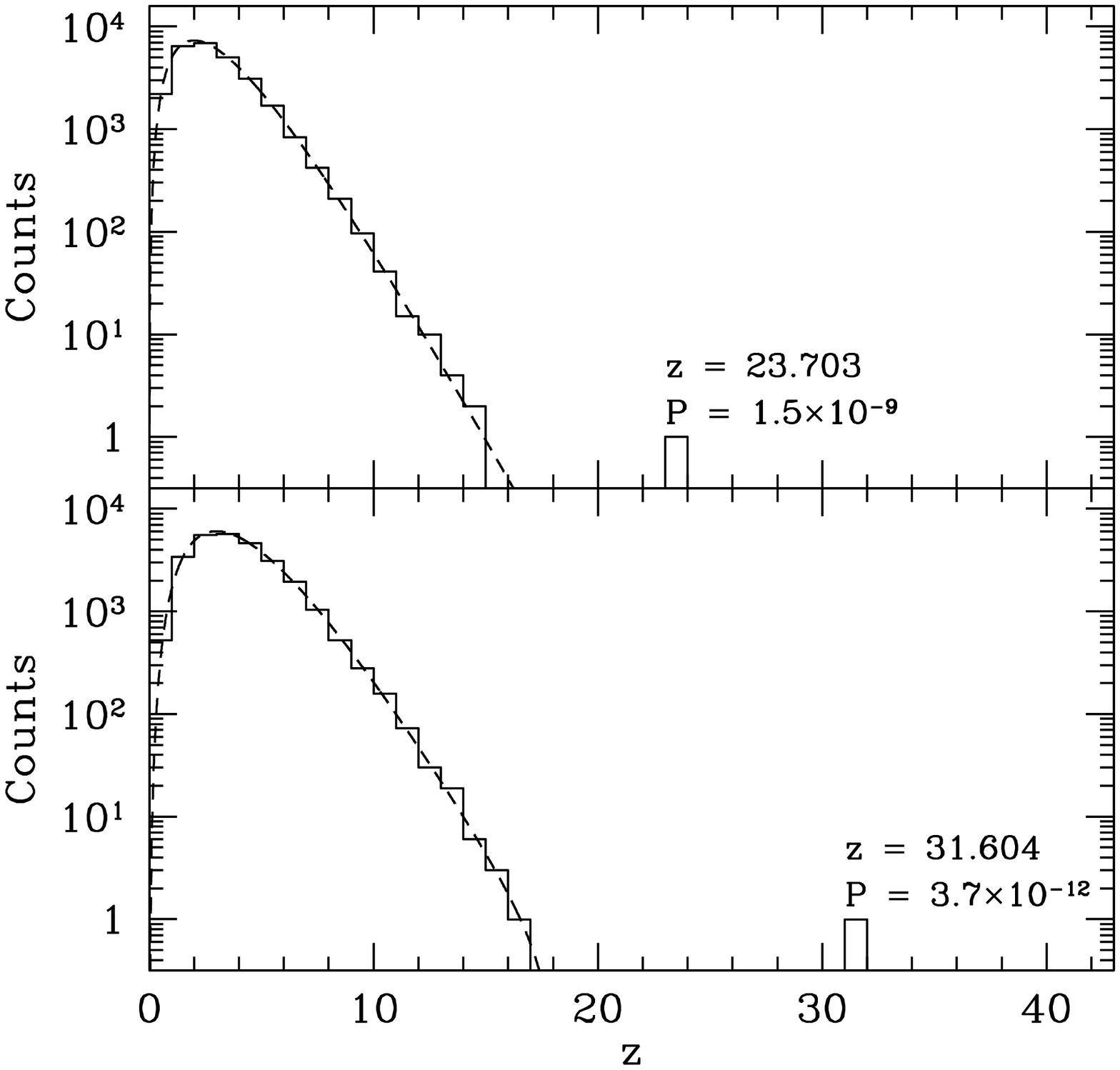}
\caption[]{Left panel: event significance calculated by the
  $\Gamma$~distribution approximation (dashed line) and the actual
  distribution (solid line), assuming $\NP = 27,000$ and $\NT =
  4$. Right panels: the results of two simulations illustrating the
  power of the rank product method for event selection. On the top
  panel, the histogram shows the parameter $z$ for $\NT = 3$ and $\NP
  = 27,000$, and the dashed line shows the true distribution given the
  null hypothesis. The rank triplet on the tail has ranks $\{10, 10,
  10\}$. The bottom plot is the same, but with $\NT = 4$, and the
  outlier arises from a rank quadruplet of $\{10, 10, 10, 10\}$.}
\label{fig:gsim}
\end{figure*}

\subsection{The Rank Product Test Statistic}

As can be seen in \fig{fig:diff}, events consistent with occultations
by KBOs will appear as one or two consecutive flux drops in all four
telescopes. Our test is thus designed to find those rank tuples where
all of the ranks are small, corresponding to a region toward the lower
left corner of the rank-rank diagram shown in \fig{fig:rr} (expanded
to $T$ dimensions, where $T$ is the number of telescopes). The
assumptions on the rank statistics and conditions placed on the raw
data outlined above allow us to calculate the significance level
$\alpha$ of various test statistics corresponding to this region.

The statistical analysis is designed to use each rank tuple $(r_{1j},
\ldots, r_{Tj})$ to perform a hypothesis test that there is an event
at time $t_j$. Each measurement $\rij$ can be used as a test statistic
for the null hypothesis of no occultation event versus the alternative
that there is an occultation, yielding a p-value given by
\begin{displaymath}
\P(R \leq \rij) = \frac{\rij}{\NP}.
\end{displaymath}
The goal is to use the tuple of $\NT$~p-values at time $t_j$ to
calculate a single test of significance. Fisher proposed that the
product of the p-values be used as a test statistic for this general
problem \citep{fisher, fisher2}. Given the product of ranks at time
$t_j$ over all telescopes $T$
\begin{displaymath}
y_j = \prod_{i=1}^{\NT}\rij,
\end{displaymath}
we define our rank product statistic as 
\begin{equation}
z_j = -\ln\left(\frac{y_j}{\NP^T}\right).
\end{equation}

Event detection based on the rank product statistic was described in
\citet{2006AN....327..814L} and \citet{2008ApJ...685L.157Z}. In the
description presented in \citet{2006AN....327..814L}, we made the
assumption that the distribution of p-values $\rij/\NP$ was uniform on
the \emph{continuous} interval $[0, 1]$ when in fact it is uniform on
the \emph{discrete} set $\{1/\NP,2/\NP, ... , 1\}$.  We found that
this assumption leads to substantial deviations from the tails of the
true distribution. For completeness, we will describe the statistical
test presented in \citet{2006AN....327..814L}, followed by a
description of the correct statistical distribution.

To calculate the probability distribution of $z_j$ under the continuous
interval assumption, we define the parameter
\begin{displaymath}
\sij = -\ln\left(\frac{\rij}{\NP}\right).
\end{displaymath}
If the p-values $\rij/\NP$ are uniform and continuous on the interval
$[0, 1]$, this quantity has the probability distribution
\begin{displaymath}
p(\sij) = e^{-\sij},
\end{displaymath}
The probability distribution of the sum of the parameters $\sij$ from
two telescopes is given by the convolution of the two distributions,
i.e.
\begin{eqnarray}
z_j &=& s_{1j} + s_{2j}\nonumber\\
p(z_j) &=& \int\limits_0^{z_j}
 dz^\prime e^{-z^\prime}e^{-(z_j - z^\prime)}\nonumber\\
&=& z_je^{-z_j}.\nonumber
\end{eqnarray}
For a total of $\NT$ telescopes, this can be generalized to the form
\begin{eqnarray}
z_j &=&\sum_{i=1}^{\NT}s_{ij},\nonumber\\
p(z_j) &=& \frac{1}{\Gamma(\NT)}z_j^{\NT - 1} e^{-z_j},
\label{eq:gparam}
\end{eqnarray}
which is simply the $\Gamma$~distribution.

Given that the distribution of p-values is discrete on the interval
$\{1/\NP,2/\NP, ... , 1\}$, the true distribution of the rank product
can be calculated using the function $\KK{n}{\NT}{\NP}$, which we
define as the number of ways to get a product of $n$ by multiplying
$\NT$ integers (number of telescopes) between 1 and $\NP$ (number of
points in the lightcurves). This function can be calculated
numerically, and we have developed a simple algorithm to calculate
$\KK{n}{\NT}{\NP}$ when $n \le \NP$ (see Appendix). Some values of
this function for $T = 4$~telescopes are shown in \tbl{tbl:K}. Note
that this function is independent of $\NP$ if $n \le \NP$.

\begin{deluxetable*}{rrrrrr}
\tablecolumns{6}
\tablewidth{0pc} 
\tabletypesize{\small}
\tablecaption{Rank quadruplets used
  to calculate $\KK{z}{\NT}{\NP}$ for $\NT=4$ and $z \le \NP$.}
\tablehead{$K(1) = 1$&$K(2) = 4$&$K(3) = 4$&$K(4) = 10$&$K(5) = 4$&$K(6) = 16$}
\startdata
1111 & 1112 & 1113 & 1114 & 1115 & 1116\\
     & 1121 & 1131 & 1141 & 1151 & 1161\\
     & 1211 & 1311 & 1411 & 1511 & 1611\\
     & 2111 & 3111 & 4111 & 5111 & 6111\\
     &      &      & 1122 &      & 1123\\
     &      &      & 1212 &      & 1132\\
     &      &      & 2112 &      & 1213\\
     &      &      & 1221 &      & 1312\\
     &      &      & 2121 &      & 2113\\
     &      &      & 2211 &      & 3112\\
     &      &      &      &      & 1231\\
     &      &      &      &      & 1321\\
     &      &      &      &      & 2131\\
     &      &      &      &      & 3121\\
     &      &      &      &      & 2311\\
     &      &      &      &      & 3211\\
\enddata
\label{tbl:K}
\end{deluxetable*}

Rather than using the function $K$ to calculate the probability
density of the rank product statistic $z$, it is simpler to calculate
the distribution as a function of the rank product $y$ as
\begin{displaymath}
p(y) = \frac{1}{\NP^\NT}\KK{y}{\NT}{\NP}.
\end{displaymath}
We thus calculate the significance, or p-value, of any candidate event as
\begin{equation}
\P(Y \le y) = \frac{1}{\NP^\NT}\sum_{i=1}^y\KK{i}{\NT}{\NP}.
\end{equation}
However, for clarity we continue to display results in terms of the
rank product statistic $z$ because candidate events are more easily
distinguished on the tail of the distribution. Given the relation
between $z$ and $y$, it clearly follows that
\begin{displaymath}
\P(Z \ge z(y)) = P(Y \le y).
\end{displaymath}
Note that the results published in \citet{2008ApJ...685L.157Z} and
\citet{fed} use the correct probability distribution based on the
discrete rank distribution.

A comparison between the discrete distribution and continuous
approximation is shown in \fig{fig:gsim}, left panel. The difference
is very significant on the tail. This is due to the fact that the most
significant rank quadruplet possible is $(1, 1, 1, 1)$, which
corresponds to $z = 40.8$, while the tail on the $\Gamma$~distribution
at $z > 40.8$ is due to values of $0 \le r < 1/\NP$, which are allowed under
the assumption of a continuous distribution of ranks.

The power of the rank product method is shown in \fig{fig:gsim} (right
panel). A three telescope (top) and a four-telescope (bottom) data
run were simulated. On the three telescope run, an event was added
with a rank triplet $\{10, 10, 10\}$, and on the four-telescope run,
an event was added with ranks $\{10, 10, 10, 10\}$. The four-telescope
event has a p-value of $3.7 \times 10^{-12}$ under the null
hypothesis, while the three-telescope event has a p-value of
$1.5\times 10^{-9}$. This simple example illustrates the value of
using multiple telescopes, in that the absence of the fourth telescope
decreases the significance of the event by more than 2,000, while
keeping the false positive rate fixed.

The rank product test statistic is based on subsets of the rank tuples
where events would plausibly be expected to be found. However, in
general, the subset of rank tuples that provides the most sensitive
detection is composed of those tuples which are most likely in the
event of an occultation.  We could imagine identifying this subset by
running an enormous simulation of occultations which produced a
probability for each of the $\NT^{\NP}$ tuples. The rejection region
for the test would then be composed of the quadruplets with largest
probabilities, the number being determined by the desired false
positive rate. Note that the rejection region might not be symmetric
in the telescopes (invariant to the telescope labels), which might be
desirable if lightcurves from some telescopes had much better signal
to noise ratios than from others.  We have not carried out such a
simulation, but a modest simulation indicates that the rejection
region determined by the rank product statistic is sufficient for the
purpose of event detection.

\subsection{False Positive Rate}
The methodology we employ is to search for an event at every time
point in every lightcurve set that the survey has collected. Hence,
the total number of hypotheses tested is
\begin{displaymath}
\Nh = \sum_l \NP(l),
\end{displaymath}
where the sum is over all lightcurve sets $l$ in the data set.

If we set a significance threshold of $\alpha$ to declare an event at
time point $t_j$ in lightcurve set $l$, and we use the same
significance threshold at all times in all light curves, then the
expected number of declared events due to chance would be
\begin{displaymath}
\alpha \times \Nh = \alpha \times \sum_l \NP(l).
\end{displaymath}
Therefore, to control false positives we must make $\alpha$ very
small. For the results published in \citet{2008ApJ...685L.157Z}, the
data set (after diagnostic cuts, see \sect{sec:cuts} for details)
comprised a total of $2.3 \times 10^9$~tuples, and the threshold used
was $\alpha = 10^{-10}$, which gives a predicted 0.23~false positive
events. To keep the false positive rate low for the larger data set
($9.0 \times 10^9$~tuples) used in \citet{fed}, we used $\alpha = 3
\times 10^{-11}$, corresponding to an expected number of 0.27 false
positives.

In all likelihood there will be at most one occultation in a
lightcurve set, and if an occultation occurred over consecutive time
points it would only be counted once.  Hence, one could consider
performing a hypothesis test over the entire light curve set rather
than at each time point by looking at the minimum of the rank product
over all time points in the light curve set:
\begin{displaymath}
\beta = \mathrm{min}_j \prod_i r_{ij}.
\end{displaymath}
If we test based on $\beta$ at level $\alpha'$, then the expected number of
false positives is
\begin{displaymath}
\sum_l \alpha' = \alpha' \times \# \{ \mbox{lightcurve sets} \}.
\end{displaymath}

The distribution of $\beta$ can be evaluated numerically, but it is
much easier to work with the rank product at every time point. Given a
constraint on the false positive rate, and given the lengths of the
series of interest, and the part of the distribution we are interested
in (the tail), it has been found \citep{nate} that that there is
little difference if we work with $\beta$ or with the rank product at
all time points; the same events will be detected.

\subsection{Detection of Multi-Point Occultation Events}
\label{sec:multipoint}
If we had an occultation from a large object in the Kuiper Belt, it
would cause a substantial flux drop for several consecutive time
points, resulting in several values of the rank product that pass the
significance threshold.  On the other hand, if the object were at 200
AU, it might cause a modest flux drop for several time points, none of
them big enough to pass the threshold.  In the latter case, it is
useful to consider functions of the data that look at neighboring time
points for detection.

Let our original time series be $ f_1 , ..., f_N $, and suppose we
form a new series by
\begin{displaymath}
a_j = b(f_{j-k}, ... , f_j , ... , f_{j+k}).
\end{displaymath}
For example, $b$ could be a moving average:
\begin{displaymath}
b(f_{j-k}, ... , f_j , ... , f_{j+k}) = \frac{1}{2k+1}(f_{j-k} + ... + f_{j+k})
\end{displaymath}

The series $a_j$ might show a larger response at the center of the
modest signal than $f_j$, leading to better detection efficiency.
Another possibility for $b$ is to take the inner product with some
signal. For example, a series of event templates could be used as the
function $b$ to search for occultation events from objects of specific
sizes and distances, which is what was done by
\citet{2009Natur.462..895S}, \citet{2009arXiv0910.5598W}, and
\citet{2008AJ....135.1039B}

Such manipulation of the data will introduce significant
autocorrelation into the lightcurves. However, if the series $f_j$ is
stationary and ergodic in mean, and if $k$ is small relative to the
length of the lightcurves, then it follows that $a_j$ will also be
stationary and ergodic in mean, so the rank product distribution will
still be satisfied. This is because the autocorrelation structure is
expected to be the same throughout the lightcurve. See \citet{nate}
for a detailed discussion.

\begin{figure*}
\plotone{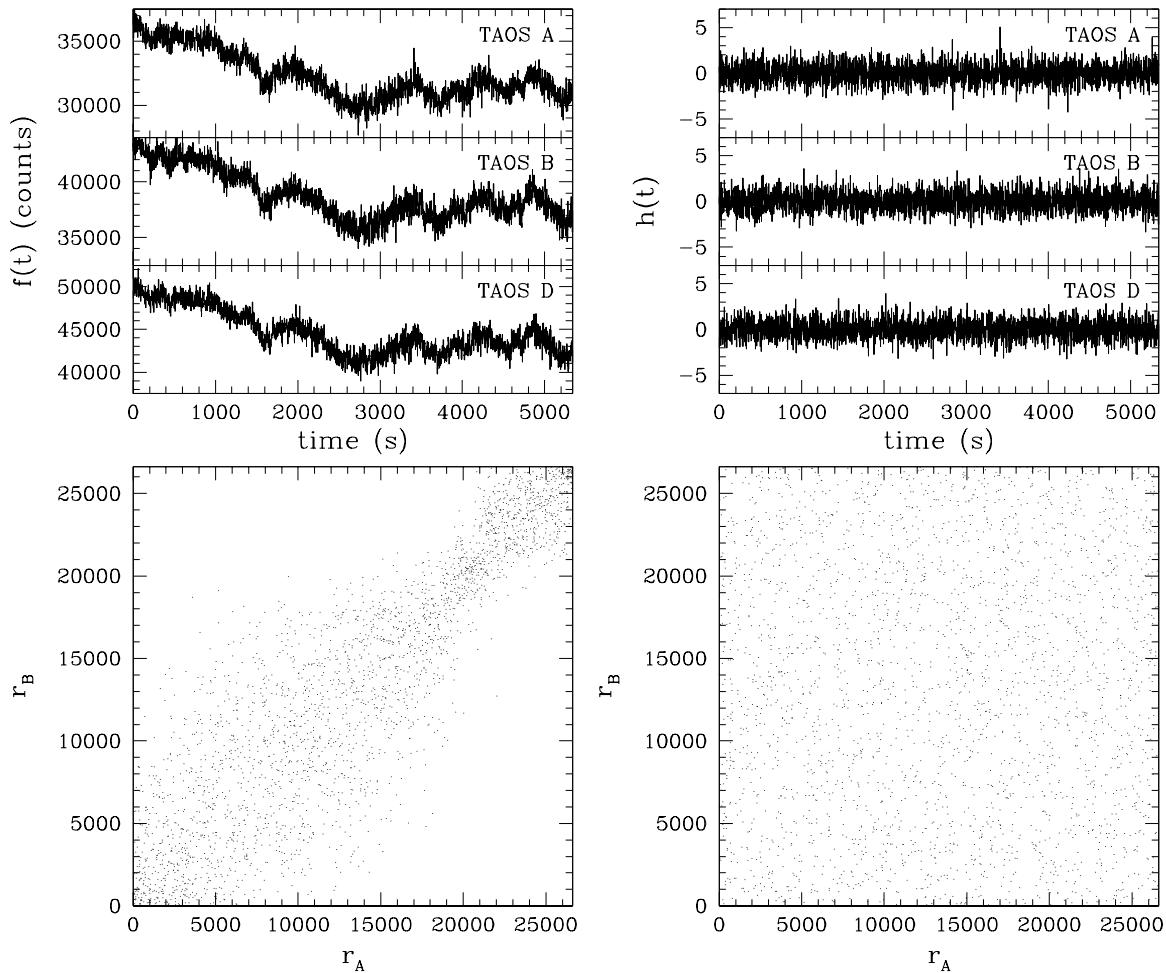}
\caption[]{Top left: an unfiltered three-telescope lightcurve set for
  a single star. Note the correlated variations in the
  lightcurves. Bottom left: rank-rank diagram for telescopes A and
  B. Top right: same lightcurve set after filtering. Bottom right:
  rank-rank diagram after filtering. All four plots are reproduced
  from \citet{2008ApJ...685L.157Z}.}
\label{fig:lc}
\end{figure*}

\section{Lightcurve Filtering}
\label{sec:filter}
As discussed above, the tests based on rank statistics are valid only
if the lightcurves from each telescope are stationary, ergodic in
mean, and independent of those from other telescopes. However, in the
actual data, significant correlations and non-stationarity are
evident, as can be seen in the top left panel \fig{fig:lc}. Trends
like those evident in the lightcurves in \fig{fig:lc} can arise due to
changing airmass and atmospheric transparency throughout the duration
of a run. The bottom left panel of \fig{fig:lc} shows the rank-rank
diagram corresponding to this lightcurve set (telescopes A and B are
shown). Under the assumption of independence, the points should be
distributed uniformly across the diagram; clearly this is not the
case.

To solve this problem, we apply a \emph{mean filter} to the
lightcurves in order to remove the slowly varying trends. Each
photometric measurement $f_j$ in a lightcurve is replaced with
\begin{displaymath}
g_j = f_j - \bar{f}_j,
\end{displaymath}
where $\bar{f}_j$ is defined as a $3\sigma$-clipped
\citep{1996A&AS..117..393B,1992ASPC...23...90D} mean taken over a
window of size $W_\mu$ which is centered on the point $f_j$.

After application of the mean filter, we found many lightcurves that
exhibit fluctuations in variance over time. In periods of higher
variance, more extreme high or low rank values are more likely, and
our assumption on the uniform distribution of ranks throughout a
lightcurve is invalid. We thus correct for changes in the variance by
applying a \emph{variance filter}, where we replace every point $g_j$
with
\begin{displaymath}
h_j = \frac{g_j}{\sigma_j},
\end{displaymath}
where the standard deviation $\sigma_j$ is calculated over a window of
size $W_\sigma$ centered on the point $g_j$, and $3\sigma$-clipping is
applied here as well.

We want to choose the window sizes to be small enough to accurately
correct for high frequency trends, but we also want them large enough
to enable accurate determination of $\bar{f}$ and $\sigma$. After
testing various window sizes, we found that $W_\mu = 33$ and $W_\sigma
= 151$ work best (the variance fluctuates much more slowly than the
mean, hence the larger window size).

We note that much work has been done in the past on removing such
trends from lightcurves, most of which involves removing correlated
trends in lightcurves from different stars in the same series of
images \citep[for example,
  see][]{2005MNRAS.356..557K,2005MNRAS.356.1466T,2009AJ....138..568B}. The
simpler approach we have adopted works well enough for our purposes,
but we are considering adopting similar techniques for future
analysis.

The top right panel of \fig{fig:lc} shows the same lightcurve set
shown in the left panel, after filtering. The trends in the mean and
variance have clearly been removed. The rank-rank diagram of the
filtered lightcurve set is shown in the bottom right panel of
\fig{fig:lc}. No dependence is evident in this diagram.

\begin{figure}
\plotone{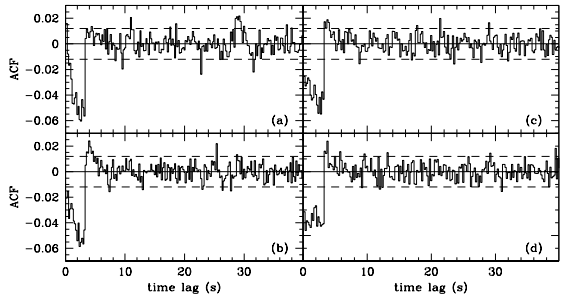}
\caption{Auto-correlation plot from four lightcurves. Panels (a), (b),
  and (c): auto-correlations plots from three filtered TAOS
  lightcurves. Panel (d): auto-correlation plot from a synthetic
  white-noise lightcurve, after application of the mean and variance
  filters. Dashed lines are the 95\% c.l. limits for what would be
  expected for randomly distributed lightcurves.}
\label{fig:acf}
\end{figure}

\fig{fig:acf} shows auto-correlation functions (ACFs) after
application of the mean and variance filters described above. Three of
the panels show ACFs of lightcurves in the TAOS data after filtering
(such as those shown in the top right panel of \fig{fig:lc}), and one
of them shows the ACF of a synthetic lightcurve of white noise, after
the same filters have been applied. The auto-correlation is
insignificant in all cases after a time lag of 6.6~seconds, which
corresponds to the window size $W_\mu$ of the mean filter. The ACF of
the simulated lightcurve demonstrates that the observed features in
the ACFs are consequences of the mean filter. The small feature in
\fig{fig:acf}a evident at time lag of 30~seconds is likely due to the
variance filter which has a window size of $W_\sigma = 151$
points. The short timescale (relative to the length of the lightcurve)
of the significant auto-correlation features is consistent with our
modeling of the filtered lightcurves as stationary and ergodic in
mean, as dependence over longer timescales invalidates our assumptions
that all possible ranks are equally likely at each time point
\citep{nate}.

\begin{figure*}
\plottwo{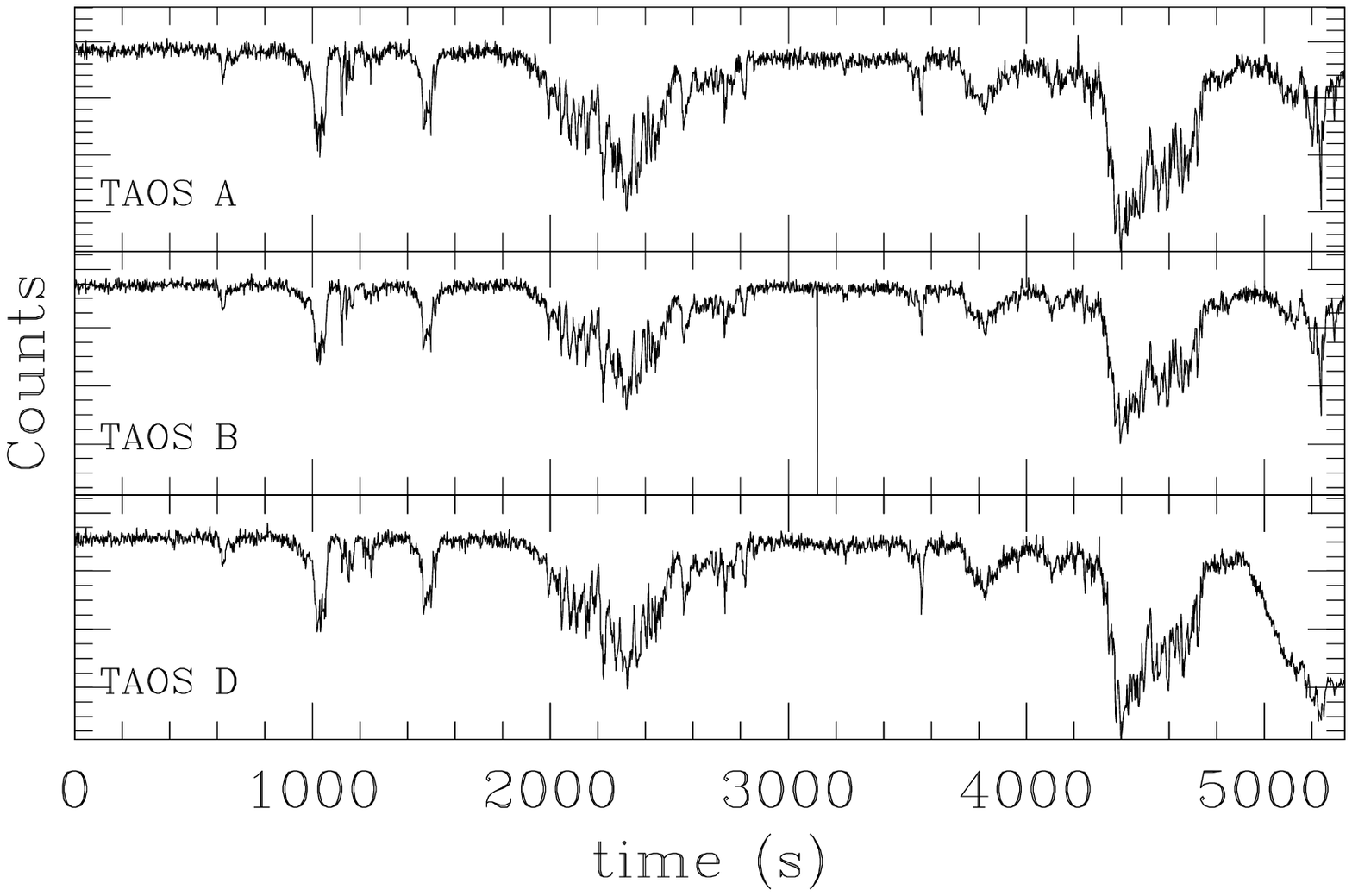}{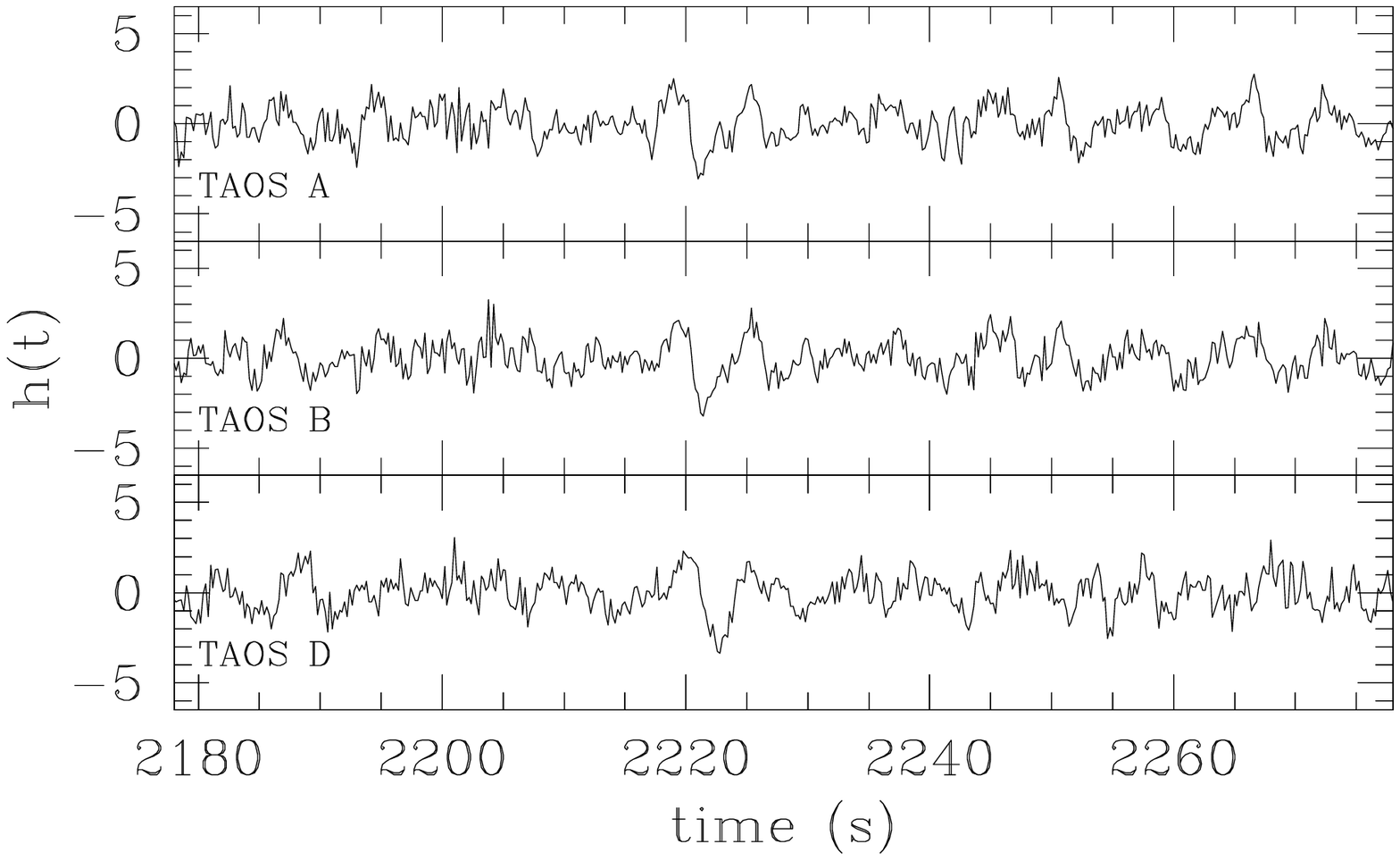}
\caption[]{Left panel: An unfiltered lightcurve set on a night with
  periods of cirrus cloud cover (during the periods of significant
  flux drops). Right panel: the same lightcurve set after filtering,
  zoomed in to a period of cloud cover. Significant correlations are
  evident. Note that the correlation is stronger between telescopes
  TAOS~A and TAOS~B, which are close together (6~m
  separation). TAOS~D, which is about 100~m away, has similar features
  but they are offset in time.}
\label{fig:badlc}
\end{figure*}

\begin{figure}
\plotone{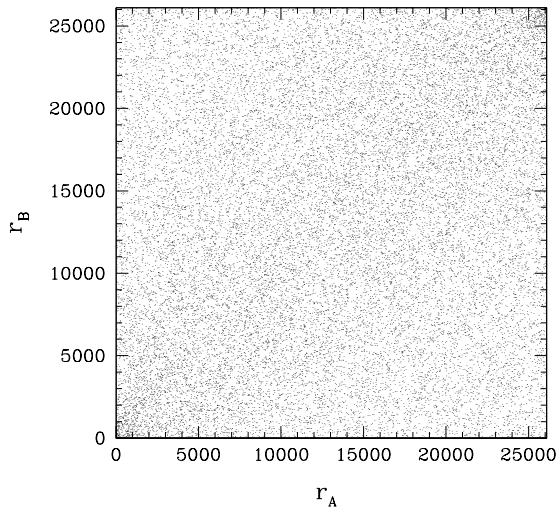}
\caption[]{Rank-rank diagram (telescopes TAOS~A and TAOS~B) of the
  lightcurve set shown in \fig{fig:badlc}. The rank pairs are not
  uniformly distributed, as there are denser than average regions in
  the lower left and upper right corners.}
\label{fig:badrk}
\end{figure}

While the filter appears to work well on the example lightcurve set
discussed above, we still need to quantify how well it actually
works. This is important because some data runs may exhibit variations
that are not adequately corrected for by the filters we apply. In
particular, data runs with extremely rapid fluctuations in the mean
(due to fast moving cirrus clouds or other phenomena) will not be
removed if the event width is small when compared with $W_\mu$. This
is illustrated in \fig{fig:badlc}, which shows a lightcurve set taken
during a night with periods of fast-moving cirrus clouds. Significant
correlations are evident in the filtered lightcurves. The
corresponding rank-rank diagram of telescopes TAOS~A and TAOS~B is
shown in \fig{fig:badrk}. Significant over-dense regions are evident
in the lower left and upper right corners of this diagram. In order
for the statistical analysis described above to be valid, such data
need to be flagged and cut from the data set before the application of
the rank product test.

We have thus developed two diagnostic tests to be applied to each
lightcurve set to assess the quality of the data after the application
of the filters. We have found that phenomena inducing correlations in
the lightcurve sets tend to affect the entire data run. Therefore,
these diagnostic tests (described in the following subsections) are
applied to entire data runs rather than individual lightcurve
sets. Data runs failing these tests are not considered for further
analysis. The tests, described in the following subsections, were used
in \citet{2008ApJ...685L.157Z}, \citet{2009AJ....138.1893W} and
\citet{fed}. An improved version of these tests has been developed for
use in future analysis runs, and these new tests will be described in
\sect{sec:newtests}.

\begin{figure*}
\plottwo{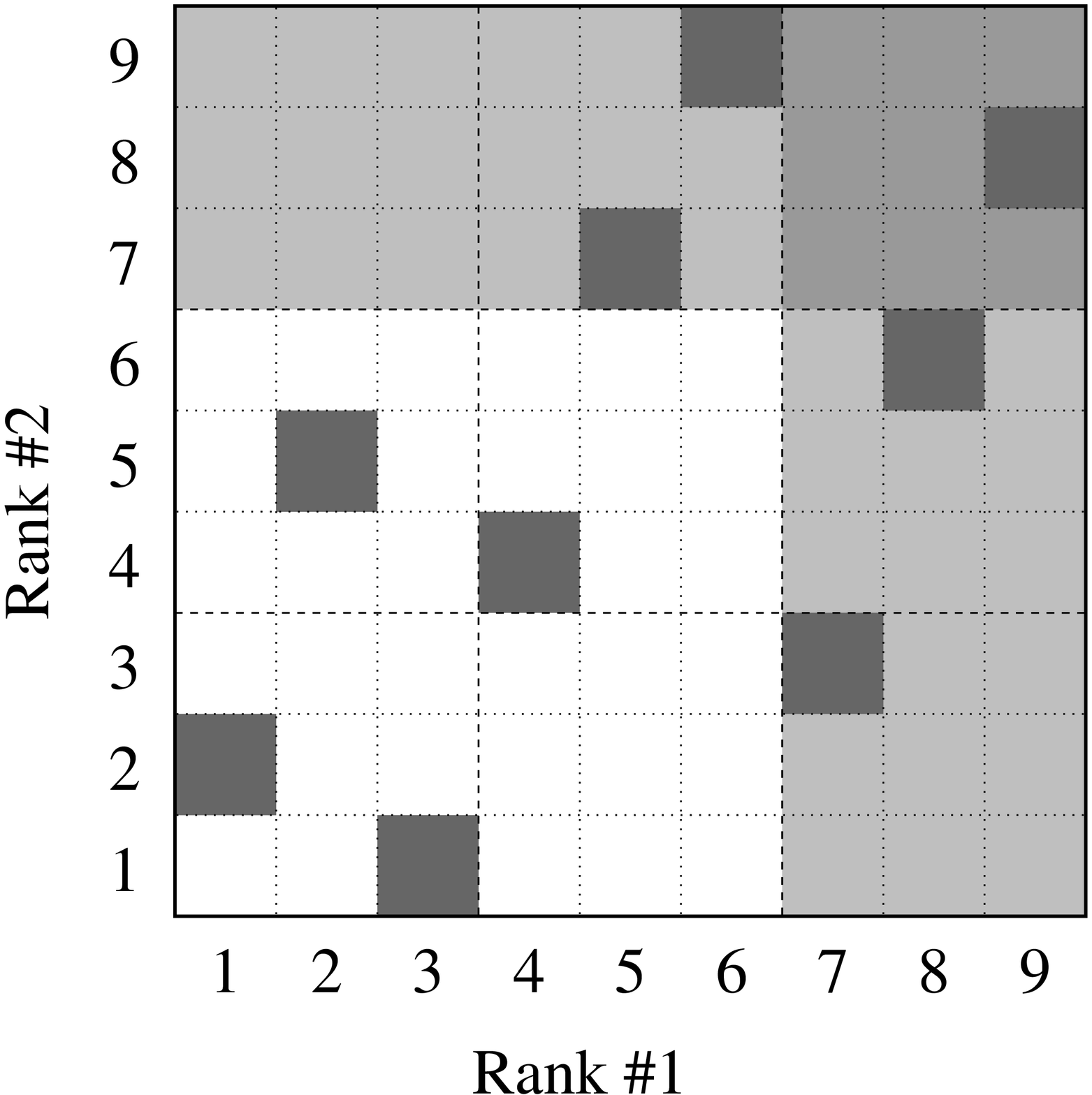}{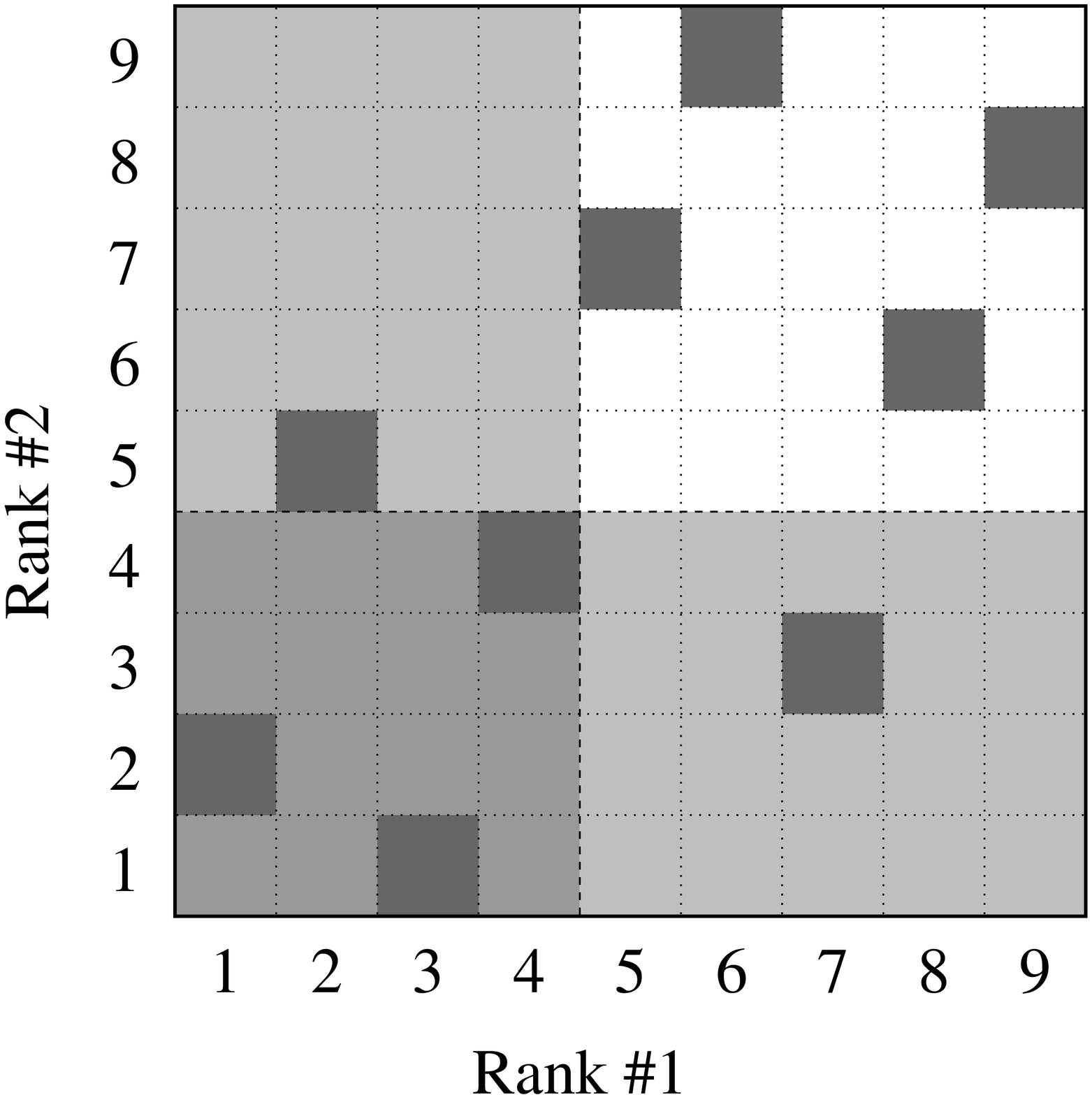}
\caption[]{Left panel: rank-rank diagram illustrating the Pearson's
  $\chi^2$ test. The rank-rank diagram is divided into a $\Ng\times
  \Ng$ grid ($\Ng = 3$ in this case), and the number of rank pairs in
  each box is tabulated and compared with the expected uniform
  distribution. Counts in the gray elements are not free parameters.
  Right panel: rank-rank diagram illustrating the hypergeometric
  test. The test counts the number of objects in the lower left corner
  (dark shaded region) of the rank-rank diagram, in this case with a
  box size of $R=4$. Note that there are four rank doublets with
  $r_{1j} \le 4$ and four rank doublets with $r_{2j} \le 4$
  (light-shaded regions) since each rank must occur exactly once in a
  lightcurve set for each telescope. In this case there are three rank
  doublets where $\rij \le 4$ for both telescopes.}
\label{fig:ziptest}
\end{figure*}

\subsection{Pearson's $\chi^2$ Statistic}

A simple test to determine if the lightcurves in a lightcurve set are
dependent is to divide the multi-telescope rank space into a grid and
count the number of rank-tuples in each grid element. This is
illustrated in the left panel of \fig{fig:ziptest}, in the case of two
telescopes where the rank-rank diagram is divided into a $\Ng \times
\Ng$~grid, where $\Ng = 3$. With $\NP=9$ and 9~grid elements, the
expected number of rank pairs in each grid element is 1.

One can then perform a Pearson's $\chi^2$ test on the number of rank
pairs in each grid element by calculating
\begin{displaymath}
\chi^{2} = \sum_{i=1}^{\Ng^{\NT}} {(O_i - \E_i)^2 \over \E_i},
\end{displaymath}
where $O_i$ is the observed number of rank-tuples in grid element $i$,
and
\begin{displaymath}
\E_i = \frac{\NP}{\Ng^{\NT}}
\end{displaymath}
is the expected number of rank-tuples in grid element $i$. (Note that
$\E_i$ may vary slightly among grid elements if $\NP$ is not an exact
multiple of $\Ng$.)

For a given data run, we expect the distribution of the Pearson's
$\chi^2$ statistic to follow the $\chi^2$ distribution, given by
\begin{equation}
p(\uC,\nu) = {1 \over 2^{\nu/2}\,\Gamma(\nu/2)}\, \uC^{(\nu/2)-1}\,e^{-\uC/2},
\end{equation}
where $\uC = \chi^2$ and $\nu$ is the number degrees of freedom.  The
derivation of $\nu$ can be illustrated by \fig{fig:ziptest}. It is
important to note that every rank must appear once and only once in
the time series for each telescope, and this constrains the value of
$\nu$. The degrees of freedom is the number of cells in the grid minus
the number of independent constraints. First, the cell counts must sum
to $\NP$, giving one constraint. Secondly, the counts in each grid row
and each grid column must sum to three, giving two constraints on the
three rows and on the three columns which are independent of each
other and of the first constraint. Thus the total degrees of freedom
are $9 - 2 - 2 - 1 = 4$.  To illustrate, note that in the left grid
column, the ranks 1, 2, and 3 must appear in telescope \#1. Therefore,
for telescope \#2, since there are two doublets in the bottom left
grid element and one doublet in the middle left grid element, there
must be zero doublets in the top left element. The gray grid elements
in the rank-rank diagram are thus not free parameters.  For an
arbitrary number of telescope $\NT$, the number of degrees of freedom
can be shown to be equal to
\begin{displaymath}
\nu = \Ng^{\NT} - \NT(\Ng-1) -1.
\end{displaymath}

\subsection{The Hypergeometric Test}

While the Pearson's $\chi^2$ test validates that the rank tuples are
spread uniformly over $\{1\ldots\NP\}^T$, it is also useful to
demonstrate that there is no bias towards rank quadruplets with all
ranks relatively low, since these are the target events in the
survey. Given a rank limit $R$, we define the variable $\uH$ as the
number of rank quadruplets with $\rij \le R$ for all telescopes $i$.
This is illustrated in the case of two telescopes in the right panel
of \fig{fig:ziptest}, where we choose $R=4$. In this figure, $\uH = 3$
is the number of rank doublets in the shaded lower left corner of the
rank-rank diagram. Note that with $R=4$ there are exactly four rank
doublets with $r_{1j} \le R$ and with $r_{2j} \le R$. The probability
distribution of the number of rank doublets with \emph{both} ranks
$\rij \le R$ is given by the hypergeometric distribution:
 \begin{equation}
\P(U = \uH) = \frac{{R \choose \uH} {\NP - R \choose R - \uH}}
  {{\NP \choose R}},
\end{equation}
where $\uH \le R$ (if $\uH > R$ then $\P = 0$).

To expand this calculation to more than two telescopes, we use the
\emph{law of total probability} to calculate
\begin{eqnarray}
\P(U_{i+1} = \uH) & = \sum\limits_{l = \uH}^{R}\,&\left[\P(U_{i+1} = \uH|U_i = l)
\right.\nonumber\\
& & \left.\times \P(U_i = l)\right],
\label{eq:ltp}
\end{eqnarray}
where $\uH$ is the number of measurements with $r \le R$ on all
telescopes 1~to~$i + 1$. The conditional probability is defined as
\begin{equation}
\P(U_{i+1} = \uH|U_{i} = l) = \frac{{R \choose \uH}{\NP-R \choose {l-\uH}}}
               {{\NP \choose l}},
\label{eq:hdist}
\end{equation}
Given that each rank must occur exactly once for each telescope,
\begin{displaymath}
\P(U_1 = l) = \delta_{lR},
\end{displaymath}
and one can thus expand \eqn{eq:ltp} to include an arbitrary number of
telescopes.

\begin{figure*}
\plottwo{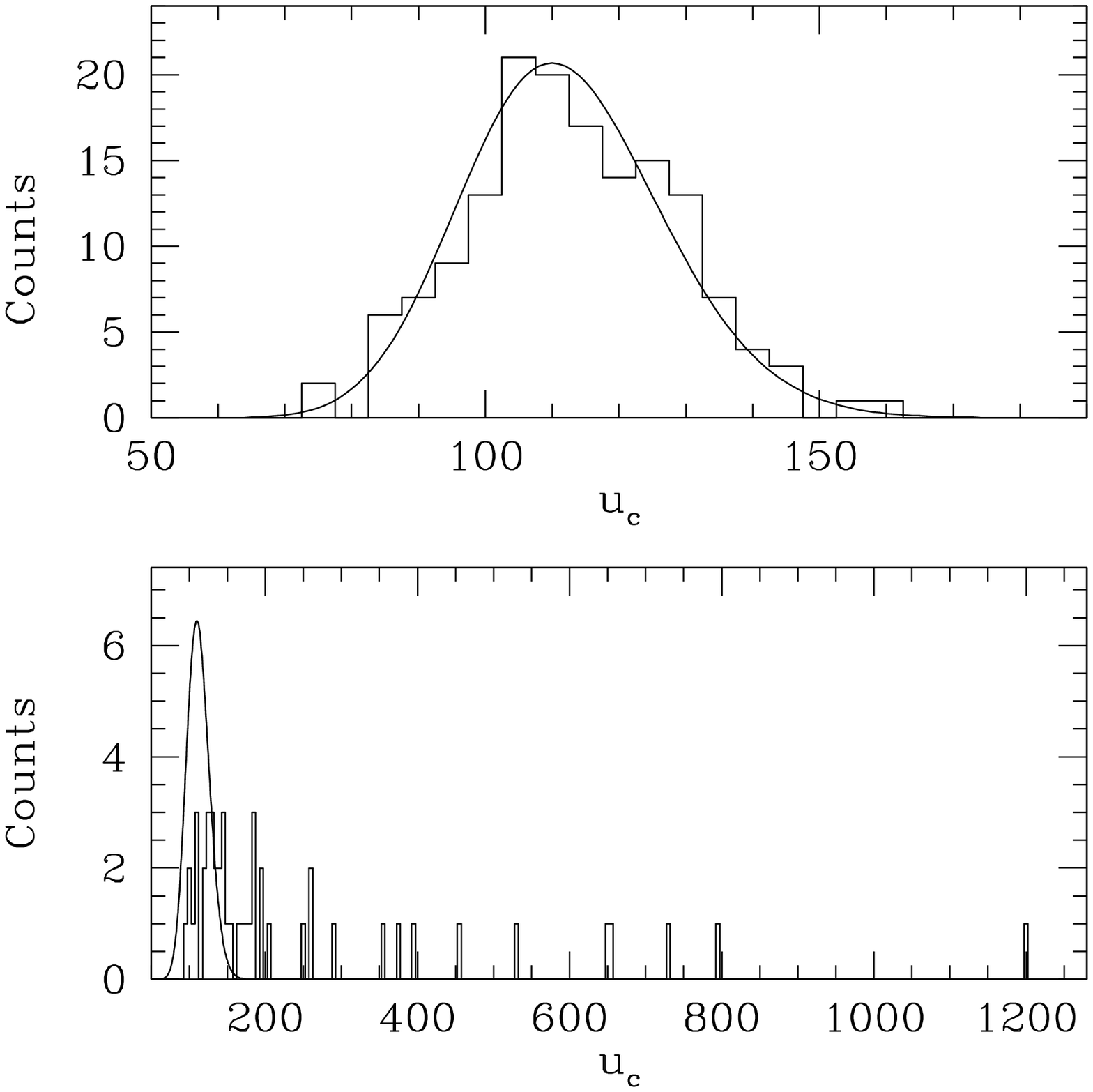}{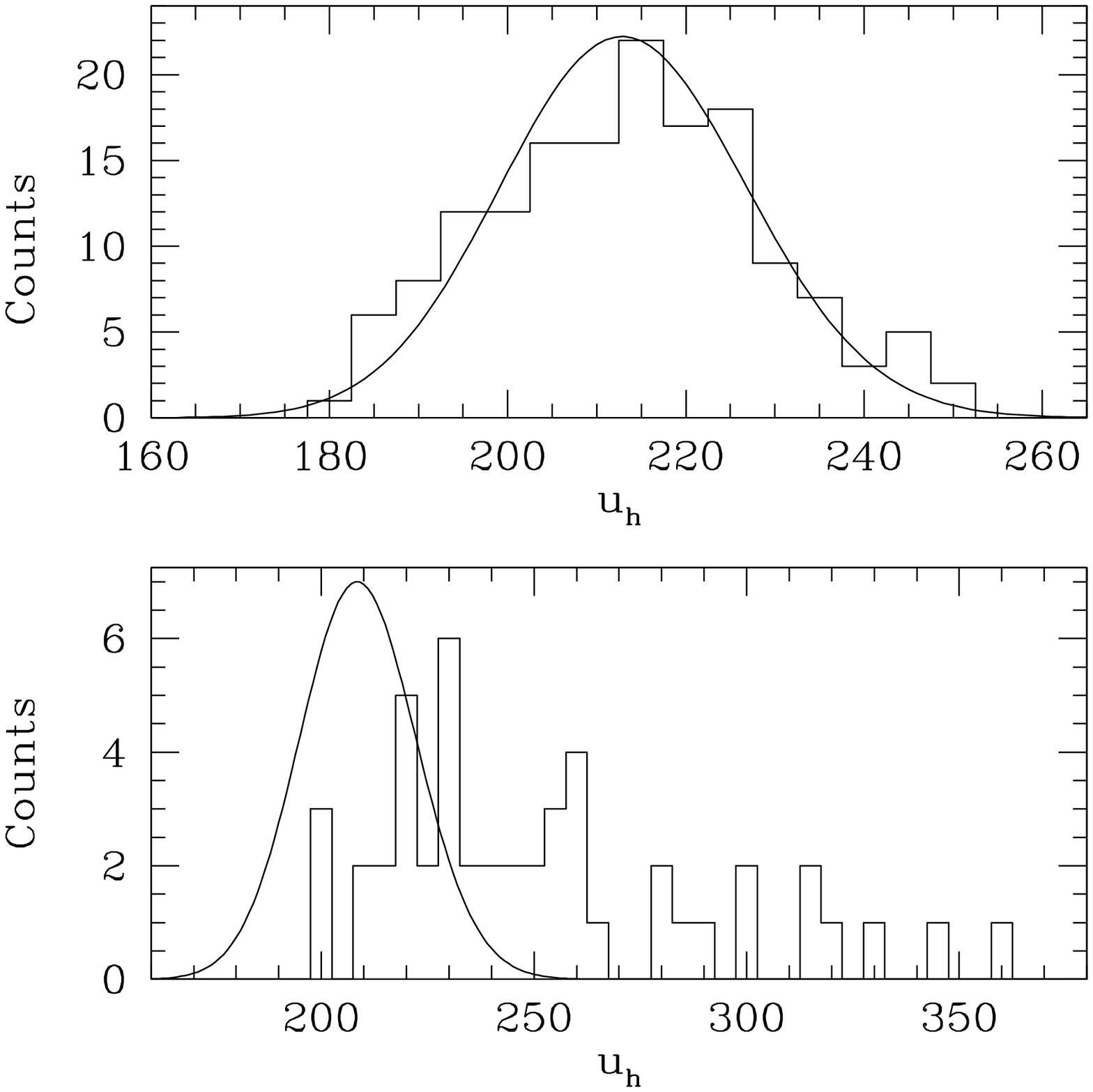}
\caption[]{Results of diagnostic tests. Histograms indicate actual
  data, and solid lines indicate theoretical distributions. Top
  panels: Pearson's $\chi^2$ test (left) and hypergeometric test
  (right) for a data run with no evident dependence between
  telescopes. Bottom panels: same as above, but for a data run with
  significant dependence between lightcurves. The lightcurve set shown
  in Figures~\ref{fig:badlc} and~\ref{fig:badrk} comes from this data
  run.}
\label{fig:badzip}
\end{figure*}

\subsection{Application of Diagnostic Statistics}
\label{sec:cuts}
To date, the TAOS project has only analyzed data sets with lightcurves
from three telescopes \citep{fed,2009AJ....138.1893W,
  2008ApJ...685L.157Z}. Therefore we only describe the application of
the diagnostic tests to three-telescope data.

For each data run, we apply both the Pearson's $\chi^2$ statistic
$\uC$ and the hypergeometric test statistic $\uH$ to each lightcurve
set. For the Pearson's $\chi^2$ test, we use a grid size of $\Ng = 5$,
which corresponds to a total of $\nu = 112$~degrees of freedom. For
the hypergeometric test, we set $R = \NP/5$, rounding to the nearest
integer. (A typical 90~minute data run will have $\NP = 27,000$, but
many runs are truncated due to bad weather.) Due to the fact that any
correlations in the data may not show up in lightcurves with low SNR
values, we perform the diagnostic tests only on those lightcurve sets
with $\mathrm{SNR} \ge 10$.  Details of the algorithm used to
calculate SNR values of our lightcurves are given in
\citet{2009PASP..121.1429Z}. To summarize, we first calculate a
5$\sigma$-clipped rolling mean similar to that which is calculated in
the mean filter, and then average the value of the rolling mean to get
the signal. We then subtract the rolling mean from the raw lightcurve,
and calculate a 5$\sigma$-clipped standard deviation of the new
lightcurve, which we use as an estimate of the noise.

Even in the case of completely independent lightcurves, random chance
will give rise to a number individual lightcurve sets with aberrant
values of the test statistic. We therefore look at the ensemble of
test statistics for each data run, and require a match to the
theoretical distributions. A set of examples is shown in
\fig{fig:badzip}. The top panels show histograms of $\uC$ and $\uH$
for all of the lightcurve sets in a data run with no evident dependence
among the lightcurves, and the bottom panels show the same data for
the pathological data run which contains the lightcurve sets shown in
Figures~\ref{fig:badlc} and~\ref{fig:badrk}. Clearly, the data shown
in the top panels match the theoretical distribution quite well, while
the data in the bottom panels do not.

\begin{figure}
\plotone{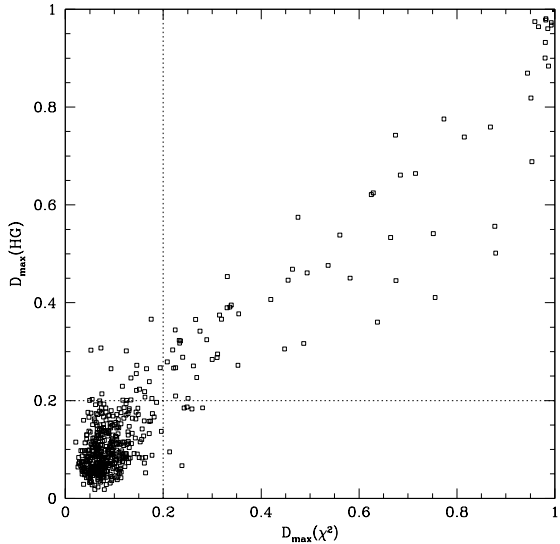}
\caption[]{Scatter plot showing $\dmax$ values for the Pearson's
  $\chi^2$ test ($x$-axis) and hypergeometric test ($y$-axis). Each
  point corresponds to a single data run. For each test statistic, we
  reject data runs with $\dmax > 0.2$ (dotted lines).}
\label{fig:dmax}
\end{figure}

To determine which data runs are to be rejected, we test the goodness
of fit of the distribution of test statistics over the lightcurve sets
in a data run to their theoretical distributions. To set a threshold,
for each data run we calculate the quantity $\dmax$, which is defined
as the absolute value of the maximum difference between the cumulative
distribution of measured test statistics and the theoretical
cumulative probability distribution. This is analogous to the
Kolmogorov-Smirnov test \citep[see][and references therein]{numrec}.
For each data run, we calculate $\dmax$ for both the Pearson's
$\chi^2$ test and the hypergeometric test. A scatter plot of these
values is shown in \fig{fig:dmax}.

Data runs that fail either of the two tests are removed from the
occultation event search. For data runs exhibiting widespread
dependence between the lightcurves from different telescopes, we
expect the measured distributions to differ significantly from the
theoretical distributions, giving rise to large values of
$\dmax$. Visual inspection of several data runs indicated that setting
a cut on $\dmax > 0.2$ allowed us to reject nearly all of the runs
where the lightcurves exhibit clear, widespread dependence.

\begin{figure}
\plotone{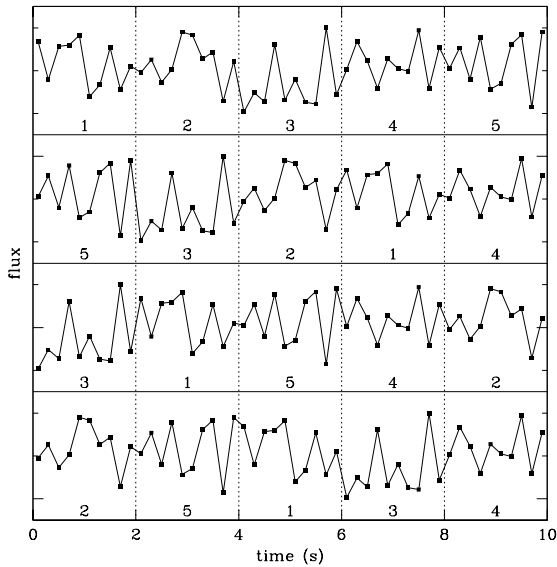}
\caption[]{Illustration of the BBS test. Top panel: original
  lightcurve, divided into five blocks of data (dotted lines). Bottom
  panels: four lightcurves with the blocks permuted randomly. Blocks
  are labeled 1~through 5~for reference.}
\label{fig:bbm}
\end{figure}

\section{Improved Diagnostic Tests}
\label{sec:newtests}
While the diagnostic tests described in the previous section are
sufficient to remove nearly all of the data runs with significant
dependence between lightcurves from different telescopes, they suffer
from some limitations which motivated us to improve the
techniques. First, the threshold $\dmax$ was chosen somewhat
arbitrarily after visual inspection of many data sets, since there is
no way to determine empirically what the optimum threshold actually
is. Second, in order for the measured statistical distributions to
match the theoretical $\chi^2$ and hypergeometric distributions, the
original time series data in the lightcurves are required to be
independent and identically distributed (\emph{i.i.d.}), which is a
stricter requirement than stationary and ergodic in mean. If some
auto-correlation structure were present in the lightcurves, the
lightcurves could still be stationary and independent from each other,
however, the test statistics would not be expected to match the
theoretical distributions. Finally, and most importantly, we would
like to apply the same tests to lightcurves when searching for
multi-point events. As discussed in \sect{sec:multipoint}, for such
event searches we would take a moving average of the lightcurve data,
or perhaps take the inner product of the lightcurve with some event
template. Such filtering will induce auto-correlations into the
lightcurves, and increase the SNR values as well. If we increase the
SNR enough, some insignificant correlations between the lightcurves
might in fact become significant in the filtered data. So it would be
useful to apply the diagnostic tests to the data runs after the
application of these filters. However, the introduction of significant
auto-correlations into the lightcurve data will more or less guarantee
that all of the data runs will fail the diagnostic tests since the
lightcurves will not be i.i.d.

We have thus developed a new technique based on the \emph{Blockwise
  Bootstrap} \citep{kunsch} method (hereinafter \emph{BBS}), which
uses both the Pearson's $\chi^2$ statistic $\uC$ and the
hypergeometric statistic $\uH$, described in the previous section, but
requires no assumptions about the theoretical distributions for either
statistic. The BBS test is implemented as follows. First, for a given
lightcurve set, we calculate both $\uC$ and $\uH$. Then we divide each
lightcurve in the lightcurve set into 100~subsets, or \emph{blocks},
of data. We then permute the blocks randomly, with each lightcurve in
the lightcurve set undergoing a different random permutation, and
recalculate the diagnostic test statistics. We repeat this step a
total of 99~times, and we are thus left with 100~statistical
measurements for each of the diagnostic tests.

This is illustrated schematically in \fig{fig:bbm}. The top panel
shows the original, unpermuted lightcurve, divided into five
blocks. The blocks are labeled 1~through 5~for clarity. The bottom four
panels show the same lightcurve with the five blocks randomly
permuted. Note that the data within each block remain unchanged.

For each diagnostic test, we have now calculated 100~different values,
one for the original lightcurve set and 99~for the randomly permuted
lightcurve sets. If we permute the blocks we still preserve the
stationary structure as long as the block size is large in comparison
to the time scale of any autocorrelation. So if the lightcurves are
independent, our 100 values are like 100 independent draws from the
same distribution. Thus, if we rank each of the series of 100 test
statistics from 1 to 100 (where a rank of one corresponds to the
largest value of $\uC$ or $\uH$, which would be the worst match to the
expected distribution), the ranks should be uniform and we can
calculate associated p-values as
\begin{equation}
\P(V < \vC) = \frac{\vC}{100}
\end{equation}
and
\begin{equation}
\P(V < \vH) = \frac{\vH}{100},
\end{equation}
where $\vC$ and $\vH$ correspond to the ranks of the test
statistics $\uC$ and $\uH$ from the unpermuted (original) light curve
sets.

\begin{figure*}
\plotone{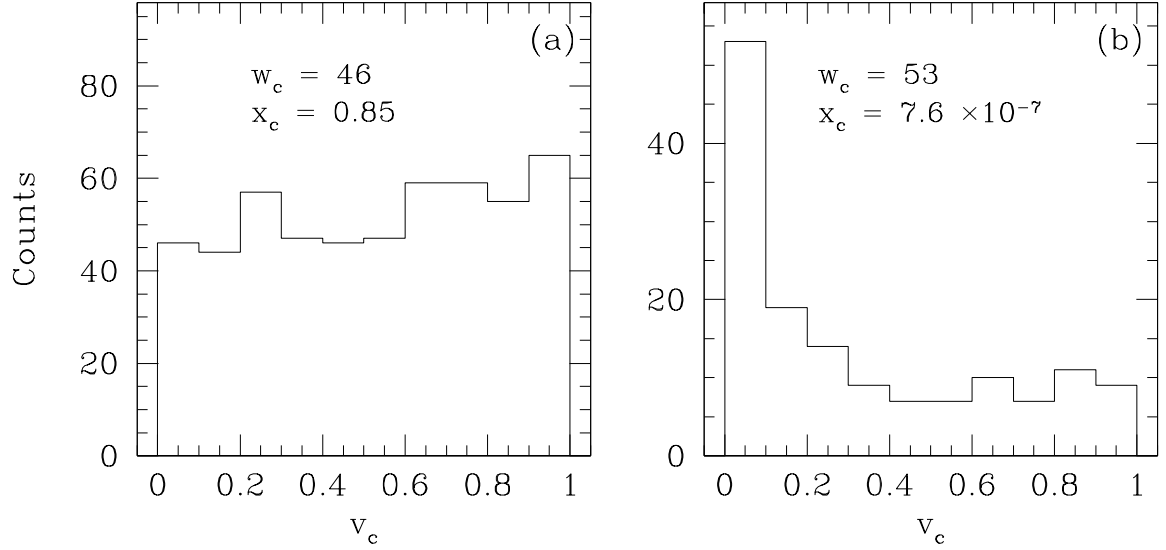}
\caption[]{Panel (a): distribution of p-values $\vC$ from the
  Pearson's $\chi^2$ test from a good data run. Panel (b): same plot
  for a rejected data run.}
\label{fig:binomial}
\end{figure*}

The BBS test is then performed on every lightcurve set in a data run
with $\mathrm{SNR} > 10$. In the case of a data run which does not
exhibit any strong dependence between the telescopes, the p-values
$\vC$ and $\vH$ should be uniformly distributed on $\{0.01, \ldots,
1\}$. However, in the case where there is significant dependence
between the telescopes, we would expect the distributions of $v$ to be
clustered at small values, since any correlation between the
telescopes would disappear when the blocks are randomly
permuted. This is illustrated in \fig{fig:binomial}, which shows a
histograms of the values $\vC$ from each lightcurve set in two
different data runs. The histogram in \fig{fig:binomial}a shows the
results from a data run with little dependence between the telescopes,
while panel \fig{fig:binomial}b shows a data run with strong
dependence.

In order to quantify the amount of dependence between the telescopes
in a data run, we define two new test statistics, $\wC$ and $\wH$,
which are defined as the number of lightcurve sets in a data run with
$\vC < \vt$ and $\vH < \vt$ respectively, where we choose $\vt =
0.1$\footnote{Tests have shown that as long as $\vt$ is relatively
  small, the exact value chosen for $\vt$ has no significant effect on
  the final results.} (This corresponds to the lowest bin in the
histograms shown in \fig{fig:binomial}). In the case of independence
between telescopes, the distributions of $\wC$ and $\wH$ follow the
binomial distribution of the form:
\begin{eqnarray}
p(\wC) &=& {L \choose \wC} \vt^{\wC} (1 - \vt)^{L - \wC},\nonumber\\
p(\wH) &=& {L \choose \wH} \vt^{\wH} (1 - \vt)^{L - \wH},\nonumber
\end{eqnarray}
where $L$ is the number of lightcurve sets with $\mathrm{SNR} > 10$ in
the data run which are used to calculate the test statistics $\uC$ and
$\uH$. Using these distributions, we can calculate two test statistics
for the entire data run, which we define as
\begin{eqnarray}
\xC(\wC) &=& \P(W > \wC),\nonumber\\
\xH(\wH) &=& \P(W > \wH).\nonumber
\end{eqnarray}
For the data run in \fig{fig:binomial}a, we thus calculate $\xC =
0.85$, while for the data run in \fig{fig:binomial}b, we have $\xC = 7.6
\times 10^{-7}$.

\begin{figure*}
\plotone{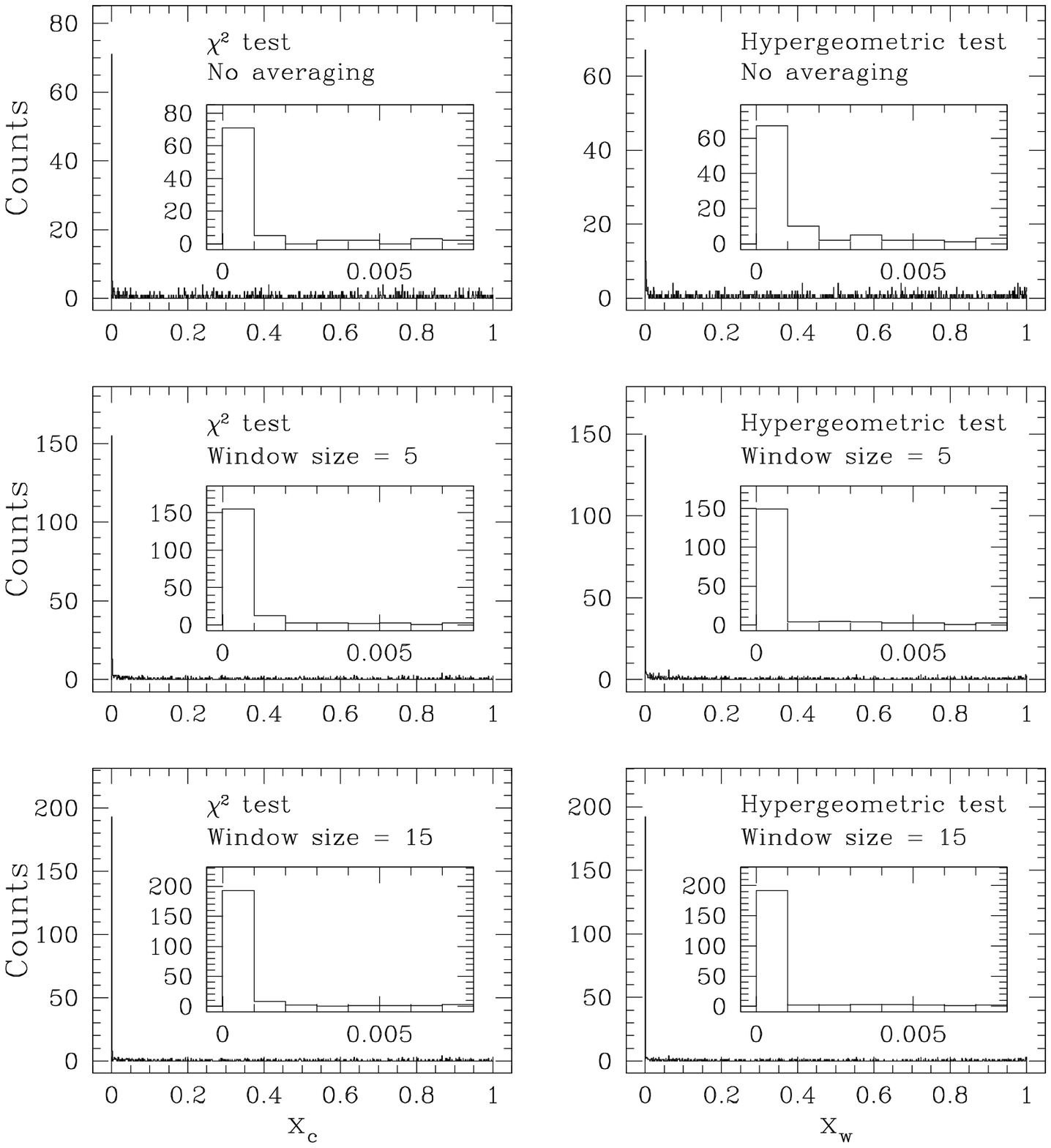}
\caption[]{Panel (a): distribution of p-values from all data runs in
  the data set described in \citet{fed} for both the $\chi^2$ and
  hypergeometric tests, with the original data and data after the
  application of moving average filters with window sizes of 5 and
  15. Inset plots are zoomed in to the lowest p-values.}
\label{fig:phist}
\end{figure*}

We can now reject a data run for significant dependence by setting
thresholds on $\xC$ and $\xH$. In the absence of any significant
dependence, the values of $\xC$ and $\xH$ should be distributed
uniformly on the interval $[0,1]$. Plots of the distributions of $\xC$
and $\xH$ statistics are shown in \fig{fig:phist}. In order to
illustrate the application of the BBS test to multi-point occultation
searches, we also show the distributions after taking moving averages
on the lightcurves with bin sizes of 5 and 15.  The histograms shown
have a bin size of 0.001, and with a total of 524 data runs we expect
a value of about 0.5 for each bin. However, note the large number of
counts in the lowest bins. These are the lightcurve sets that show
dependence between telescopes. Note that some of the histograms show a
slight excess in the second bin as well.  By rejecting all data runs
that appear in the first two bins, we are clearly rejecting nearly all
of the data runs exhibiting widespread dependence between the
telescopes.  Note that we only expect a total of one data run in the
first two bins from random chance. Furthermore, note that the larger
the bin size on the moving average, the more data runs that are
rejected. This is because of low level correlations that are
insignificant in the unbinned data, but become significant in the
binned data due to the increased SNR of the binned lightcurves.

The BBS test is clearly a superior method to the simple comparison of
the test statistics to their theoretical distributions. It is very
clear where the thresholds on $\xC$ and $\xH$ should be, the test will
not reject data runs where the lightcurves are stationary but not
i.i.d., and the tests are capable of robustly rejecting data runs when
performing searches for multi-point occultation events.

\section{Conclusion}
We have developed a technique to search for extremely rare coincident
events in voluminous multivariate (multi-telescope) time series
data. Using rank statistics, this technique enables robust
determination of event significance and false positive rate,
\emph{independent of the underlying noise distributions in the time
  series data.} This method has been used to search for rare
occultation events by KBOs in over 500 data runs comprising a total of
nearly 370,000 lightcurve sets. Furthermore, we have developed a
method to test for widespread dependence between lightcurves in a data
run, which allows us to reject runs with inherent characteristics
which could possibly give rise to a larger false positive rates. We
note that while the method described in this paper is sufficient for
the calculation of the rate of false positive events that arise due to
random statistical chance, it is not capable of estimating the
background event rate due to systematic errors in the TAOS photometry
\citep{2009PASP..121.1429Z}. For example, tracking errors or moving
objects in the images could give rise to false detections in the data
set. A description of how such background events are handled is
described in \cite{fed}.

\acknowledgements This work was supported in part by the thematic
research program AS-88-TP-A02 at Academia Sinica. NKC's work was
supported in part by the National Science Foundation under grant
DMS-0636667. Work at the Harvard College Observatory was supported in
part by the National Science Foundation under grant AST-0501681 and by
NASA under grant NNG04G113G.  The work at National Central University
was supported by the grant NSC 96-2112-M-008-024-MY3. JAR's work was
supported in part by the National Science Foundation under grant
AST-00507254. SLM's work was performed under the auspices of the
U.S. Department of Energy by Lawrence Livermore National Laboratory in
part under Contract W-7405-Eng-48 and by Stanford Linear Accelerator
Center under Contract DE-AC02-76SF00515. YIB acknowledges the support
of National Research Foundation of Korea through Grant
2009-0075376. KHC's work was performed under the auspices of the
U.S. Department of Energy by Lawrence Livermore National Laboratory in
part under Contract W-7405-Eng-48 and in part under Contract
DE-AC52-07NA27344.

\appendix
\section{Evaluation of the K function}

Here we present an algorithm to evaluate $\KK{x}{k}{n}$, the number of
ways to get a product of $x$ by multiplying $k$ integers between 1 and
$n$, which is applicable when $x \leq n$.

Note that $\KK{1}{k}{n} = 1$.  For $x > 1$, consider the prime
decomposition of $x$ where the $p$'s are unique primes and $d$ is their
degree so that:

\begin{displaymath}
x = p_1^{d_1} \times p_2^{d_2} \times \ldots \times p_m^{d_m}.
\end{displaymath}
where $m$ is the total number of prime factors of $x$.

We claim that

\begin{displaymath}
\KK{x}{k}{n} = \prod_{i=1}^m {d_i + k - 1 \choose{k-1}}
\end{displaymath}

\subsection{Proof}

Suppose $A_1 \times A_2 \times \ldots \times A_k = x$ and take prime
decompositions of each number:

\begin{figure}
\plotone{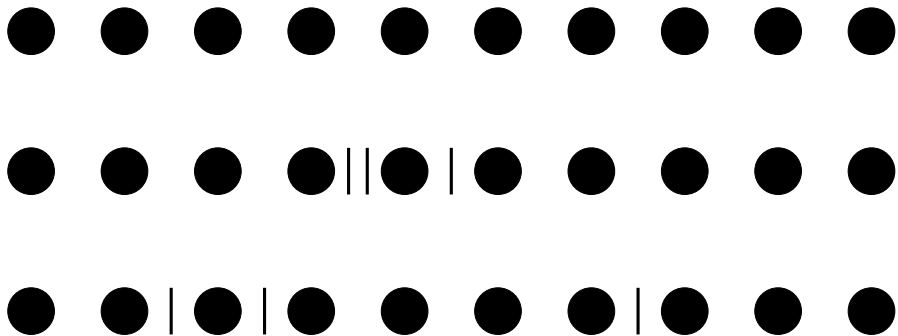}
\caption{Schematic illustrating the calculation of the function
  $S(d;k)$. In this case, $d=10$, as indicated by the top row of
  10~dots. The bars in the bottom two rows indicate possible ways to split
  the ten dots into four addends. The second row indicates the
  tuple $(4, 0, 1, 5)$, and the third row indicates the
  tuple of $(2, 1, 4, 3)$.}
\label{fig:sfun}
\end{figure}

\begin{eqnarray}
A_1 &=& p_1^{d_{1,1}} \times p_2^{d_{1,2}} \times \ldots \times p_m^{d_{1,m}}
\nonumber\\
A_2 &=& p_1^{d_{2,1}} \times p_2^{d_{2,2}} \times \ldots \times
p_m^{d_{2,m}}
\nonumber\\
&&\vdots \quad \quad  \quad \quad \quad \vdots \quad \quad \quad
\quad \vdots \nonumber\\
A_k &=& p_1^{d_{k,1}} \times p_2^{d_{k,2}} \times \ldots \times
p_m^{d_{k,m}}
\nonumber
\end{eqnarray}
Note that
\begin{displaymath}
\sum_{i=1}^k d_{i,j} = d_j \quad \forall j.
\end{displaymath}
Hence,
\begin{displaymath}
\KK{x}{k}{n} = \prod_{i=1}^m \SSS{d_i}{k},
\end{displaymath}
where
\begin{displaymath}
\SSS{d}{k} = \# \left\{(a_1,\ldots,a_k) \in \{0,..,d\}^k \ \mathrm{s.t.}
\sum_{i=1}^k a_i = d \right\}
\end{displaymath}
is the number of ways to get a sum of $d$ by adding $k$ integers where
$0 \leq k \leq d$. The calculation of the function $S$ is best
illustrated with an example. Consider the case of $d = 10$ and $k =
4$. If we illustrate the sum $d=10$ with 10~dots in the top row of
\fig{fig:sfun}, the function $S$ is simply the number of ways to
divide the dots into 4~groups (using the bars shown). For example, the
second row of \fig{fig:sfun} corresponds to a tuple of $(4, 0, 1, 5)$,
while the third row corresponds to a tuple of $(2, 1, 4, 3)$. So the
number of possible 4-tuples is the number of ways to choose 3~bar
locations in a total $10 + 3 = 13$ possibilities. This gives
\begin{displaymath}
\SSS{d}{k} = {d + k - 1 \choose{k-1} }.
\end{displaymath}

For example, to calculate $\KK{6}{4}{n}$ where $n \ge 6$, we note
that $6 = 2 \times 3$ is the product of two primes to the first power,
and in the 4 telescope case it is equal to
\begin{eqnarray}
\KK{6}{4}{n} &=& {1 + 4 - 1 \choose{4-1}}^2\nonumber\\
&=& {4 \choose 3}^2\nonumber\\
&=&4^2\nonumber\\
&=& 16,\nonumber
\end{eqnarray}
in agreement with \tbl{tbl:K}. Note that, as mentioned above, this
formulation is only valid if $x \le n$. For example, $\KK{6}{4}{5} =
12$, since any of the rank tuples with a rank value of 6 would be
impossible in a lightcurve set containing only 5~points.

As a second example, for the case of $360 = 2^3 \times 3^2 \times 5$,
we have three primes with degrees 3, 2, and 1. We thus have (for $n
\ge 360$)
\begin{eqnarray}
\KK{360}{4}{n} &=&
{3+4-1\choose 4-1}{2+4-1\choose 4-1}{1+4-1\choose 4-1}\nonumber\\
&=& {6 \choose 3}{5 \choose 3}{4 \choose 3}\nonumber\\
&=& 20 \times 10 \times 4\nonumber\\
&=& 800.\nonumber
\end{eqnarray}

\bibliographystyle{apj}
\bibliography{apj-jour,ms}

\end{document}